\documentclass[useAMS,usenatbib,
]{mnras}
\usepackage{graphicx}
\usepackage[section]{placeins}

\usepackage{times} 
\usepackage{amsmath}
\usepackage{amssymb}
\usepackage{tikz}
\usepackage{xcolor}

\def\icarus{Icarus}

\def\aj{Astron.\ J. }
\def\apj{Astrophys.\ J. }
\def\apjl{Astrophys.\ J. }
\def\apjs{Astrophys.\ J.\ (Supp.) }
\def\aap{Astron.\ Astrophys. }

\def\mnras{Mon.\ Not.\ R.\ Astron.\ Soc. }

\def\planss{Plan.\ Space Sci. }

\begin{document}

\title[Disturbing function for {nearly} polar orbits]{The disturbing function for polar Centaurs and transneptunian objects}
\author[F. Namouni and M. H. M. Morais]{F. Namouni$^{1}$\thanks{E-mail:
namouni@obs-nice.fr (FN) ; helena.morais@rc.unesp.br (MHMM)} and  M. H. M. Morais
$^{2}$\footnotemark[1]\\
$^{1}$Universit\'e C\^ote d'Azur, CNRS, Observatoire de la C\^ote d'Azur, CS 34229, 06304 Nice, France\\
$^{2}$Universidade Estadual Paulista (UNESP), Instituto de Geoci\^encias e Ci\^encias Exatas, Av. 24-A, 1515 13506-900 Rio Claro, SP, Brazil}

\date{Accepted 2017 July 6. Received 2017 July 6; in original form 2017 April 12}

\maketitle

\begin{abstract}
The {classical} disturbing function of the three-body problem is based on an expansion of the gravitational interaction in the vicinity of nearly coplanar orbits. Consequently, it is not suitable for the identification and study of resonances of the Centaurs and transneptunian objects on nearly polar orbits with the solar system planets. Here, we provide a series expansion algorithm of the gravitational interaction in the vicinity of polar orbits and produce explicitly the disturbing function to fourth order in eccentricity and inclination cosine. The properties of the polar  series differ significantly from those of the {classical} disturbing function: the polar series {can} model any resonance as the expansion order is not related to the resonance order. The powers of eccentricity and inclination  of the force amplitude of a $p$:$q$ resonance do not depend on the value of the resonance order $|p-q|$ but only on its parity. Thus all even resonance order eccentricity amplitudes are $\propto e^2$ and odd ones $\propto e$ to lowest order in eccentricity $e$. With the new findings on the structure of the polar disturbing function and the possible resonant critical arguments, we illustrate the dynamics of the polar resonances 1:3, 3:1, 2:9 and 7:9 where transneptunian object 471325 {could} currently be  locked. 
\end{abstract}

\begin{keywords}
celestial mechanics--comets: general--Kuiper belt: general--minor planets, asteroids: general -- Oort Cloud.
\end{keywords}

\section{Introduction}\label{sec1}
 The increasing detections of Centaurs and transneptunian objects (TNOs) on nearly polar orbits \citep{Gladmanetal09,Chen16} raises the question of their origin and relationship to the solar system planets. Among the dynamical processes that govern the evolution of such objects are mean motion resonances. In this context, it was shown recently, through intensive numerical simulations, that mean motion resonances are  efficient  at polar orbit capture \citep{NamouniMorais15,NamouniMorais17}.  It is therefore important to have a  thorough understanding of the processes of {resonance crossing and capture} for nearly polar Centaurs and TNOs so that we {can} identify the pathways  that led to such orbits and ultimately uncover their origin. 

Identifying a mean motion resonance for a Centaur or a TNO with a solar system planet is a fundamentally simple task. One has to search for angle combinations of the form  $\phi=q\lambda-p\lambda^\prime-k\varpi+(p-q+k)\Omega$ that {can} be stationary  or oscillating around the equilibrium {defined by the condition} $\dot \phi=0$. In the previous expression, $\lambda$ and $\lambda^\prime$ are respectively  the  mean longitudes of the object and the planet, $\varpi$ and $\Omega$ are respectively  the object's longitudes of perihelion and ascending node, and $p$, $q$ and $k$ are integer coefficients. In the angle combination $\phi$, we ignored the planet's perihelion and node because the solar {system} planets{'} eccentricities and inclinations with respect to the invariable plane are small. As the number of integer combinations is infinite, one usually seeks and checks only the strongest resonances: those with a force amplitude that implies a sizable resonance width within which to capture the Centaur or TNO. This choice is as reasonable as it is rewarding provided that one remembers that the force amplitudes associated with a candidate resonance $\phi$ are obtained from the {classical} disturbing function which is an expansion in powers of eccentricity and inclination of the planet-object gravitational interaction for nearly {circular} and coplanar orbits. Thus our intuition regarding the  angle combination $\phi$ and its dynamical suitability for resonance is based on the assumption of near-coplanarity. It is the object of this work to remedy this shortcoming in the dynamical analysis of polar Centaurs and TNOs by deriving a disturbing function for nearly polar orbits and studying the properties of its force amplitudes.
 
The history of the {classical} disturbing function is intertwined with that of  celestial mechanics. For a historical perspective we refer the reader to \cite{BC61}. For the purposes of this work, we note that the disturbing function of the three-body problem takes two different forms. The first form is a power series in terms of eccentricity $e$ and $\sin^2(I/2)$ where $I$ is the inclination. This  form therefore assumes that the object's orbit does not depart significantly from prograde coplanar motion. It is used widely to study the formation and dynamics of planetary systems, the formation and evolution of planetary rings and the formation and resonance capture of planetary satellite systems \citep{EllisMurray00,ssdbook}. The second form is a power expansion in terms of the  ratio $\alpha=a/a^\prime$ where $a$ and $a^\prime$ are the semi-major axes of the object and  planet respectively. This form is used mainly to study the dynamics of artificial and natural planetary satellites that  have large inclinations as they are influenced by the distant Sun or the Moon. It was recently revisited for the study of the secular evolution of hierarchical planetary systems  \citep{Laskar10}. In the case of Centaurs and TNOs, this second form is not particularly useful  as such objects {can} be quite close to the planets' orbits but unlike satellites they revolve around the Sun not the planet. Reasonable order expansions with respect to $\alpha$  {can} not model the dynamics when the semi-major axes' ratio does not satisfy  $\alpha\ll1$. We will therefore seek a disturbing function that is not expanded with respect to $\alpha$ but is written as a power series of eccentricity and some function of the inclination that vanishes if the object's orbit is exactly polar. We find in Section \ref{sec2} that the natural function is simply $\cos I$. The {classical} disturbing function and its zero reference inclination {can} also be transformed into a disturbing function for nearly coplanar retrograde orbits, that is with $180^\circ$ reference inclination, to study the dynamics of retrograde resonances \citep{MoraisNamouni13b}. The retrograde disturbing function helped {to} identify the first Centaurs and Damocloids in retrograde resonance with Jupiter and Saturn \citep{MoraisNamouni13a}.

 The plan of the paper is as follows. In Section \ref{sec2}, we write down the explicit steps of the literal expansion of the gravitational interaction of a  planet with a particle on  a nearly polar orbit in powers of eccentricity and inclination cosine. The reader who is not interested in the details of the expansion algorithm {can} skip this part and find the resulting disturbing function in Section \ref{sec3}. The properties of the polar disturbing function are compared to those of the {classical} disturbing function of nearly coplanar prograde orbits as well as that of nearly coplanar retrograde orbits in Section \ref{sec4}. The validity domain of the polar disturbing function is linked to secular evolution and discussed in Section \ref{sec5}.  Examples of polar resonances are found in Section \ref{sec6}.  Section \ref{sec7} contains concluding remarks. 
  
\section{Literal expansion for nearly polar orbits}\label{sec2}

Consider a test particle {of negligible mass} that moves under the gravitational influence of the sun  of mass $M_{\star}$ and a planet of mass $m^\prime\ll M_{\star}$. The motion of $m^\prime$ with respect to $M_{\star}$ is a circular orbit of  radius $a^\prime$ and longitude angle $\lambda^\prime$.  The reference plane is defined by the sun-planet orbit. The test particle's osculating Keplerian orbit with respect to $M_{\star}$  has semi-major axis $a$, eccentricity $e$, inclination $I$, true anomaly $f$, argument of pericentre $\omega$, and  longitude of ascending node $\Omega$.  After normalizing all distances to $a^\prime$, the  disturbing function  reads:
\begin{equation}
R= {G\,m^\prime}{a^{\prime-1}} (\Delta^{-1}-r\,\cos{\psi})\equiv {G\,m^\prime}{a^{\prime-1}}\bar{R},
\end{equation}
where $r=\alpha (1-e^2)/(1+e \cos f)$ is the orbital radius of the particle and $\alpha=a/a^\prime$ is the particle's normalized semimajor axis that {can} be larger or smaller than unity,  $\psi$ is the angle between the radius vectors of  the planet and the test particle,
$\Delta^2=1+r^2-2\,r\,\cos{\psi}$ is the planet-particle relative distance and
\begin{equation}
\cos\psi=\cos(\Omega-\lambda^\prime)\cos(f+\omega )-\sin(\Omega-\lambda^\prime)\sin(f+\omega )\cos I.  \label{cospsi}
\end{equation}
The first term of $\bar{R}$  is the gravitational force from the mass $m^\prime$ on the test particle also known as the direct perturbation that we shall denote $\bar{R}_d$. The second term, that we denote $\bar{R}_i$, is the indirect perturbation that comes from the reflex motion of the star under the influence of the mass $m^\prime$ as the standard coordinate system is chosen to be  centered on the star. In the following we use the notations and steps of the literal expansions for  nearly coplanar prograde orbits  \citep{ssdbook} and nearly coplanar retrograde orbits {by} \cite{MoraisNamouni13a} so that the reader is able to see the similarities and differences of the three expansions.

The {classical} series of the disturbing function is expanded in powers of $e$ and $\sin^2(I/2)$ which is adequate for nearly coplanar prograde motion since $\sin^2(I/2)$ vanishes for $I=0$.  The {classical} series {can} also be used for nearly coplanar retrograde orbits after having applied the procedure devised by  \citep{MoraisNamouni13a} that allows one to get retrograde resonant terms from prograde ones. In essence, the retrograde series is an expansion in terms of $e$ and $\cos^2(I/2)$ where the latter vanishes for $I=180^\circ$.  However, neither the prograde series nor its retrograde counterpart {can} be used for polar orbits.  Instead, inspection of the expressions of $\cos\psi$ (\ref{cospsi}) and $\Delta$  reveals that a polar expansion has to be done with respect to $e$ and $\cos I$ that vanishes for $I=90^\circ$. We therefore write  
\begin{equation}\Delta^2=1+r^2-2 r
\cos(\Omega-\lambda^\prime)\cos(f+\omega )-2r\Psi,\end{equation} 
where  $\Psi$ is:
\begin{equation}
\Psi = -\sin(\Omega-\lambda^\prime)\sin(f+\omega )\cos I.\label{PSI}
\end{equation}
Expanding the direct perturbation term $\Delta^{-1}$ in the vicinity of $\Psi=0$,  we have:
\begin{eqnarray}
\Delta^{-1} &=& \sum_{i=0}^{\infty} \frac{(2 i) !}{2^i(i !)^2} (r\Psi)^{i}  {\Delta_{0}^{-(2\,i+1)}},\label{Deltaexp}
\end{eqnarray}
where $\Delta_0^2=1+r^2-2\,r\,\cos(\Omega-\lambda^\prime)\cos(f+\omega )$. Defining  $\epsilon=r/\alpha-1=\mathcal{O}(e)$ and expanding $\Delta_{0}^{-(2\,i+1)}$ around $\epsilon=0$, we get:
\begin{equation}
{\Delta_{0}^{-(2i+1)}} =
\left( 1+\sum_{k=1}^{\infty}\frac{\epsilon^k\,\alpha^k}{k!}\,\,\frac{d^k}{d \alpha^k} \right) {\rho^{-(2i+1)}}, \label{Delta0}
\end{equation}
\begin{equation}
{\rho^{-(2i+1)}}=   [1+\alpha^2-2\,\alpha\,\cos(\Omega-\lambda^\prime)\cos(f+\omega )]^{-(i+1/2)}. \label{rho2i}
\end{equation}
{The validity of the expansion with respect to zero eccentricity will be discussed in Section \ref{sec5} where we examine the secular coupling of eccentricity and inclination. In particular, a maximum value of eccentricity will be determined for the polar disturbing function of fourth order. }

The next step {in the literal expansion} is to develop the function $\rho^{-(2i+1)}$ into a {two-dimensional }Fourier series with respect to the angles $f+\omega$ and $\Omega-\lambda^\prime$ as follows:
\begin{equation}
 {\rho^{-(2i+1)}}=\! \! \! \! \! \! \! \sum_{
{\scriptsize\begin{array}{c}
-\infty<j,k<\infty\\ 
j+k \ {\rm even}
\end{array}}} \! \! \! \frac{1}{4}b_{i+1/2}^{jk}(\alpha) \cos[k(f+\omega )+j(\Omega-\lambda^\prime)], \label{RHO}
\end{equation}
\begin{equation}
b_{s}^{jk}(\alpha) = \frac{1}{\pi^2}\int_0^{2\pi} \int_0^{2\pi} \frac{\cos(j u+kv) \ du\, dv}{(1+\alpha^2-2\alpha \cos u \cos v)^s}, 
\end{equation}
where $b_{s}^{jk}(\alpha)$ are two-dimensional Laplace coefficients. The series (\ref{RHO}) is summed over even $j+k$ owing to the invariance of the function $\rho^{2i+1}$ (\ref{rho2i}) with respect to the variable change $(f+\omega+\pi,\Omega-\lambda^\prime+\pi)$ that makes $b_{s}^{jk}=0$ if $j+k$ is odd. This invariance {can} be interpreted as a {combined} change of reference for $\Omega$ and $\omega$ to the descending node. We remark that the presence of the two-dimensional Laplace coefficients is related to the fact that the two angles $f+\omega$ and $\Omega-\lambda^\prime$ enter the expression of $ {\rho^{-(2i+1)}}$ independently in contrast to the expansions of nearly coplanar orbits where {these} angles enter as the sum  $f+\omega\pm(\Omega-\lambda^\prime)$ where the $\pm$ signs are for prograde and retrograde orbits respectively \citep{MoraisNamouni13a}. The two-dimensional Laplace coefficients also satisfy the following relations:
\begin{eqnarray}
 b_{s}^{jk}&=& b_{s}^{kj}= b_{s}^{(-j)k}= b_{s}^{j(-k)}= b_{s}^{(-j)(-k)},\label{Lap}\\
D\, b_{s}^{jk}&=&\frac{s}{2} \left(b_{s+1}^{(j+1)(k+1)}+b_{s+1}^{(j+1)(k-1)}+b_{s+1}^{(j-1)(k+1)}+\right. \nonumber \\
&&\left.+b_{s+1}^{(j-1)(k-1)}\right)- 2 \alpha s b_{s+1}^{jk},\\
D^n \, b_{s}^{jk}&=&\frac{s}{2} \left(D^{n-1} \,b_{s+1}^{(j+1)(k+1)}+D^{n-1} \,b_{s+1}^{(j+1)(k-1)}+\right. \nonumber \\
&&\left.+D^{n-1} \,b_{s+1}^{(j-1)(k+1)}+D^{n-1} \,b_{s+1}^{(j-1)(k-1)}\right)\nonumber \\ &&- 2 \alpha s D^{n-1} \,b_{s+1}^{jk}- 2(n-1)sD^{n-2}\,b_{s+1}^{jk},
\end{eqnarray}
where the operator $D=d/d\alpha$. Substituting the series (\ref{RHO}) into the expression (\ref{Delta0}) and the latter into the expansion (\ref{Deltaexp}), the direct part of the perturbation is  written as the following series: 
\begin{equation}
\bar{R}_d =\!\! \!\!\!\!\!\!\!\!\!\!\!\!\!\sum_{
{\scriptsize\begin{array}{c}
0\leq i,l<\infty\\ 
-\infty<j,k<\infty\\ 
j+k \ {\rm even}
\end{array}}}\!\!\!\!\!\!\!\!\!\!\!\! \!\!\!\frac{1}{l!} \frac{(2 i) !}{2^{i+2}(i !)^2} (  r \Psi )^{i} \epsilon^l A_{i,j,k,l}\cos[k(f+\omega )+j(\Omega-\lambda^\prime)],\label{RX}
\end{equation}
where 
\begin{equation}
A_{i,j,k,l}=\alpha^l D^l  b_{i+1/2}^{jk},
\end{equation}
satisfy the same symmetry of the two-dimensional Laplace coefficients in that $A_{i,j,k,l}=0$ if $j+k$ is odd. It now remains {that we express the terms}  $\Psi$ and $r$ as functions of  the mean longitude and the longitude of pericentre. This can be achieved using the {classical} elliptic expansions:
\begin{eqnarray}
\sin f&=& 2 (1-e^2)^\frac{1}{2}\sum_{s=1}^\infty \frac{d}{sde}J_s(se)\sin [s (\lambda-\varpi)],\nonumber\\
\cos f&=& -e +{2(1-e^2)}e^{-1}\sum_{s=1}^\infty J_s(se)\cos [s (\lambda-\varpi)],\nonumber\\
\frac{r}{\alpha}&=& 1+\frac{e^2}{2}-2e\sum_{s=1}^\infty \frac{d}{s^2de}J_s(se)\cos [s (\lambda-\varpi)]. \label{elliptic}
\end{eqnarray}
Upon substituting the expressions (\ref{elliptic}) into the direct part of the perturbation (\ref{RX}) and truncating it to order $N$ in eccentricity and inclination cosine, we obtain a series that is not quite ready for use for a general $p$:$q$ resonance. Indeed, the series contains  cosine terms whose seemingly unrelated arguments must be transformed to make them represent the same $p$:$q$ resonance. For example, among the various terms that appear to second order in eccentricity and zero order in inclination cosine, there are: 
\begin{eqnarray}
T_1\!\!\!\!\!\!&=&\!\!\!\!\!\!-\frac{e^2}{4} k^2 A_{0,j,k,0} \cos [k \lambda-j \lambda^\prime+(j-k) \Omega ], \\
   T_2\!\!\!\!\!\!&=&\!\!\!\!\!\!\frac{e^2}{32} A_{0,j,k,2} \cos [(k-2)\lambda-j \lambda^\prime+(j-k)   \Omega +2 \varpi].
\end{eqnarray}
Both these terms {can} be made to correspond to the same resonance $p$:$q$ by choosing $j=p$, $k=q$ for $T_1$ and $j=p$ and $k=q+2$ for $T_2$. The transformed terms become: 
\begin{eqnarray}
T_1\!\!\!\!\!\!&=&\!\!\!\!\!\!-\frac{e^2}{4}  q^2 A_{0,p,q,0} \cos [q \lambda-p \lambda^\prime+(p-q) \Omega ], \\
T_2\!\!\!\!\!\!&=&\!\!\!\!\!\!\frac{e^2}{32}  A_{0,p,q+2,2} \cos [q\lambda-p \lambda^\prime+(p-q-2)   \Omega +2 \varpi].
\end{eqnarray}
Furthermore for $p\neq0$ and $q\neq 0$, one needs to account for the resonant terms that are generated by $T_1$ and $T_2$ under the change $p\rightarrow -p$ and $q\rightarrow -q$ as the series (\ref{RX}) is summed over positive as well as negative $k$ and $j$. This transformation produces two new terms, $T_3$ and $T_4$, that correspond to the same resonance:
\begin{eqnarray}
T_3\!\!\!\!\!\!&=&\!\!\!\!\!\!-\frac{e^2}{4}  q^2 A_{0,p,q,0} \cos [q \lambda-p \lambda^\prime+(p-q) \Omega ]=T_1, \\
T_4\!\!\!\!\!\!&=&\!\!\!\!\!\!\frac{e^2}{32}  A_{0,p,q-2,2} \cos [q\lambda-p \lambda^\prime+(p-q+2)   \Omega -2 \varpi].
\end{eqnarray}
In the indices of $A_{0,p,q,0}$ and $A_{0,p,q-2,2}$, we use the properties  (\ref{Lap}) of the two-dimensional Laplace coefficients. The secular terms {can} be obtained by setting $p=0$ and $q=0$ in $T_1$ and $T_2$ but not in $T_3$ and $T_4$ because the same term would be counted twice. Another way of seeing this is that  $(0,0)$ is a fixed point of the transformation  ($p\rightarrow -p$, $q\rightarrow -q$).

For the indirect part of the disturbing function, $\bar{R}_i$, one requires only the use of the elliptic expansions (\ref{elliptic}) to transform true anomalies into mean anomalies  and {perform} the eccentricity expansion.  The resulting expressions of the direct and indirect parts of the polar disturbing function are given in the next Section. The secular part of the disturbing function is given in Section \ref{sec5}.

\section{Disturbing function of nearly polar orbits}\label{sec3}
\subsection{Direct part}
For a general resonance  $p$:$q$ and an expansion {of} order $N$ in eccentricity and inclination cosine, the steps described in the previous section show that the direct part of the disturbing function is given as:
\begin{eqnarray}
\bar{R}_d&=&\!\!\!\!\!\!\!\!\sum_{
{\scriptsize\begin{array}{c}
-N\leq k\leq N \\
|k|\leq m\leq N\\
0\leq n\leq N\\ 
m+n=N 
\end{array}}}\!\!\!\!\!\!\!\!
c^{k}_{mn}(p,q,\alpha)\, e^m \cos^n I\ \cos (\phi -k \omega), \label{RY}\\
\phi&=&q \lambda -p \lambda^\prime +(p-q) \Omega, \nonumber
\end{eqnarray}
and $\omega=\varpi-\Omega$ is the argument of pericentre. The force coefficients $c^{k}_{mn}(p,q,\alpha)$ are given explicitly for the fourth order series $N=4$ and $k=0$, 1, 2, 3, and 4 in Tables \ref{t1} to \ref{t5}.  For negative $k$, the force coefficients {can} be obtained from the identity  $c^{-k}_{mn}(p,q,\alpha)=c^{k}_{mn}(-p,-q,\alpha)$ --see for instance the {example} {of} $T_4$ above. Examination of the force coefficients shows the additional relationship $c^{-k}_{mn}(p,q,\alpha)=c^{k}_{mn}(p,-q,\alpha)$. {We note that the force coefficients are applicable to inner and outer perturbers as the semi-major axis {ratio} can be smaller or larger than unity \citep{Williams69}.  }

The force coefficients $c^{k}_{mn}(p,q,\alpha)$ have a further important property related to the resonance order $o_r=|p-q|$.  Examination of  $c^{k}_{mn}(p,q,\alpha)$'s dependence on $A_{i,j,k,l}$ and recalling that $A_{i,j,k,l}=0$ when $j+k$ is odd (or equivalently $j-k$ is odd\footnote{The integers $j+k$ and $j-k$ have the same parity.}) show that for an odd resonance order $o_r$, $c^{k}_{mn}(p,q,\alpha)=0$ when $k$ is even. Similarly for an even resonance order $o_r$, $c^{k}_{mn}(p,q,\alpha)=0$ when $k$ is odd. This property guarantees that the integer coefficient of the longitude of ascending node, $\Omega$, that reads $p-q+k$ is always even. 

To illustrate this property, we shall consider the even order inner 5:1 resonance and the odd order outer 2:9 resonance and write down the corresponding series to second order in eccentricity and inclination cosine. Using Tables \ref{t1} and \ref{t3} for the 5:1 resonance, we get: 
\begin{eqnarray}
\bar{R}_d^{5:1}&=& [c^0_{00}(5,1,\alpha)+c^0_{01}(5,1,\alpha)\cos I+c^0_{20}(5,1,\alpha)e^2+ \nonumber \\
&&+c^0_{02}(5,1,\alpha)\cos^2I]  \cos(\lambda-5\lambda^\prime+4\Omega)\nonumber \\
&& +c^2_{20}(5,1,\alpha)e^2\cos(\lambda-5\lambda^\prime+6\Omega-2\varpi)  \label{5:1}\\
&& +c^2_{20}(5,-1,\alpha)e^2\cos(\lambda-5\lambda^\prime+2\Omega+2\varpi). \nonumber
\end{eqnarray}
The resonant terms $\cos( \lambda-5\lambda^\prime+5\Omega-\varpi)$ and $\cos(5\lambda-\lambda^\prime+3\Omega+\varpi)$ whose force amplitudes are proportional to  $e$ and to $e \cos I$ {can} not appear as the corresponding force coefficients all vanish because $o_r=4$ is even. They are written as:
\begin{eqnarray}
c_{10}^{1}(5,1,\alpha) &=& -\frac{1}{4}A_{0, 5, 0, 0} =0,\\ 
c_{10}^{1}(5,-1,\alpha) &=&-\frac{1}{4} (4A_{0, 5, 2, 0} + A_{0, 5, 2, 1})=0, \\
c_{11}^{1} (5,1,\alpha) &=& \frac{\alpha}{16}  [-(A_{1, 4, 1, 0} - A_{1, 4, 1, 0}+ A_{1, 6, 1, 0}\nonumber\\&&- A_{1, 6, 1, 0}) - A_{1, 4, 1, 1}  + A_{1, 4, 1, 1} \nonumber\\&& +  A_{1, 6, 1, 1}  - A_{1, 6, 1, 1}]=0,\\
c_{11}^{1} (5,-1,\alpha) &=& \frac{\alpha}{16}  [-5 (A_{1, 4, 3, 0} - A_{1, 4, 1, 0}+ A_{1, 6, 1, 0}\nonumber\\&&- A_{1, 6, 3, 0}) - A_{1, 4, 3, 1}  + A_{1, 4, 1, 1} \nonumber\\&& +  A_{1, 6, 3, 1}  - A_{1, 6, 1, 1}]=0,
\end{eqnarray}
where we have used the two-dimensional Laplace coefficient relations (\ref{Lap}) and the property $A_{i,j,k,l}=0$ when $j+k$ is odd. 
We remark that unlike the {classical} disturbing function for nearly coplanar orbits, the resonance order does not appear in the powers of eccentricity and inclination cosine of the force amplitudes. Moreover, such terms  as (\ref{5:1}) would not exist {in} the {classical} disturbing function as the lowest order pure eccentricity term would be proportional  to $e^4$.  To dispel doubt on the existence of an unknown symmetry that would make the force coefficients of (\ref{5:1}) vanish,  we list their non-zero numerical values for nominal resonance $\alpha=0.341995$: 
$c^0_{00}(5,1)=0.00069676$, 
$c^0_{01}(5,1)=0.000586162$, 
$c^0_{20}(5,1)=0.00432225$, 
$c^0_{02}(5,1)=-0.00111703$, 
$c^2_{20}(5,1)=0.00242241$ and 
$c^2_{20}(5,-1)=0.00397014$. 
Furthermore, in Section \ref{sec6} we show examples of capture  in  high order resonances for low values of the integer $k$ that are smaller than {the resonance order}  $o_r$ (see also the next paragraph). This more general  fundamental difference between the two disturbing functions of nearly coplanar and nearly polar orbits will be discussed in Section \ref{sec4}.

The next example is given by the second order in eccentricity and inclination cosine series of the 2:9 resonance  that is free of second order eccentricity terms  because $o_r=7$ is odd. The corresponding expressions are obtained from Table \ref{t2} and read:
\begin{eqnarray}
\bar{R}_d^{2:9}&=&  e[c^1_{10}(2,9,\alpha)+c^1_{11}(2,9,\alpha)\cos I]\times \nonumber \\ &&\cos(9\lambda-2\lambda^\prime-6\Omega-\varpi) + \label{2:9}\\
 &&e[c^1_{10}(2,-9,\alpha)+c^1_{11}(2,-9,\alpha)\cos I]\times \nonumber \\ &&\cos(9\lambda-2\lambda^\prime-8\Omega+\varpi). \nonumber
 \end{eqnarray}
 The values of the various force coefficients evaluated at nominal resonance, $\alpha=2.72568$, are: $c^1_{10}(2,9)=0.000227527$, $c^1_{11}(2,9)=0.000149595$, $c^1_{10}(2,-9)=8.68079\times 10^{-6}$ and  $c^1_{11}(2,-9)=15.6653\times 10^{-6}$. 
Similarly to the previous example, the terms in (\ref{2:9}) would not exist in the {classical} disturbing function as the lowest order pure eccentricity term would be proportional  to $e^7$.   Using the numerical integration of the full equations of motion, examples of the 2:9 resonance are given in Section \ref{sec6} where libration is shown to occur in the $e$-resonant term ($k=1$) but also in higher {order} terms such as $e^3$ ($k=3$) and $e^5$ ($k=5$) that like  (\ref{2:9}) would not exist in the {classical} disturbing function of nearly coplanar orbits for the seventh order 2:9 resonance. 

\subsection{Indirect part}
The arguments and force amplitudes up to and including fourth order of the disturbing function's indirect part for nearly polar orbits are given in Table \ref{tind}. The terms present in the expansion concern only resonances of the type 1:$n$ with $0\leq n\leq 5$. {These terms therefore concern only perturbers located inside the object's orbit}.

\section{Comparing the disturbing functions of nearly coplanar orbits and nearly polar orbits}\label{sec4}
The first main difference between the disturbing functions of nearly coplanar orbits and nearly polar orbits is the fact that the expansion order is not related to the resonance orders. To understand this difference, recall that a literal expansion of the {classical} disturbing function (of nearly coplanar orbits) to order $N$ in eccentricity and inclination produces cosine terms that represent at most resonances of order $N$. For instance, the fourth order series of \cite{ssdbook} applied to a  particle perturbed by  a planet on a nearly coplanar prograde circular orbit produces only the cosine terms: $\cos[j (\lambda - \lambda^\prime)+f_0(\varpi, \Omega) ]$, $\cos[j \lambda - (j-1)\lambda^\prime+f_1(\varpi, \Omega) ]$, $\cos[j \lambda - (j-2)\lambda^\prime+f_2(\varpi, \Omega) ]$, $\cos[j \lambda - (j-3)\lambda^\prime+f_3(\varpi, \Omega) ]$, and  $\cos[j \lambda - (j-4)\lambda^\prime+f_4(\varpi, \Omega) ]$, where the functions $f_i$ represent the correct combinations of the longitudes of pericentre and ascending node that we do not reproduce explicitly to avoid cumbersome notation. This shows that the possible resonances are of order zero to four but no more. The literal expansion of the disturbing function of nearly polar orbits produces cosine terms for any type of resonance $p$:$q$ with no restriction on the resonance order $o_r=|p-q|$. A close inspection of the expansion shows that this property is related to the presence of the two independent angles $f+\omega$ and $\Omega-\lambda^\prime$ that require the use of two-dimensional Laplace coefficients unlike the {classical} disturbing function that makes use of one-dimensional Laplace coefficients. Therefore{\bf,} whereas the nearly polar disturbing function of  order 4 {can} be used for the 2:9 resonance discussed in the previous section to study the $e$-terms associated with  $k=1$  and $k=3$, a literal expansion of order 7 of the {classical} disturbing 
function is required to get the first possible resonant term. This property motivated \cite{EllisMurray00} to come up with  an algorithm that produces the  force amplitude of a given cosine term of any resonance order without having to expand the {classical} disturbing function literally. 

The second main difference between the two disturbing functions is the fact that the powers of eccentricity and inclination cosine in the force amplitudes of the polar disturbing function  are independent of the value of the resonance order $o_r$ (except its parity discussed in the next paragraph). In the {classical} disturbing function, the lowest order eccentricity and inclination power of the  force amplitude of a given cosine term is $o_r$. Indeed for any resonance of order $o_r$,  the force amplitude of the cosine  term  is proportional to $e^{o_r-2k} \sin(I/2)^{2k}$ to the lowest order in $e$ and $I$ where the integer $k$ satisfies $0\leq 2k\leq o_r$. The examples of the 5:1 and 2:9 resonance in the previous section showed that their force amplitudes  to lowest order in $e$ and $\cos I$ were:  {affine  with respect to $\cos I$ as $c_{00}^0+c^0_{01}\cos I$ and quadratic in $e$ through the terms, $c^0_{20}e^2$, $c^2_{20}e^2$ and $c^{-2}_{20}e^2$ for  5:1 (Equation \ref{5:1}).  To lowest order in $e$ and $\cos I$, the 2:9 force amplitude is linear in $e$ through the terms $c^1_{10} e$ and $c^{-1}_{10}$ (Equation \ref{2:9}).} In the {classical} disturbing function, if we seek a linear dependence in eccentricity for 2:9, we must carry along the inclination to the sixth power.  The only inclination-free force amplitude is proportional to $e^7$.

{When the algorithm of \citep{MoraisNamouni13a} is applied to the classical disturbing function (of prograde orbits) to produce a series for nearly coplanar retrograde orbits, the force amplitude of a $p$:$q$ retrograde resonance to lowest order in $e$ and  $\cos(I/2)$ is proportional  to $e^{|p+q|-2k} \cos(I/2)^{2k}$ where $0\leq 2k \leq |p+q|$.  This gives even to a first order  resonance (i.e. $|p-q|=1$),  the dynamical structure of a high order resonance. For instance, the planar 1:2 resonance is equivalent to the third order 1:4 resonance \citep{MoraisGiuppone12}.  Therefore unlike the polar case, retrograde force amplitudes  involve a retrograde resonance order defined as $\bar{o}_r=|p+q|$. }

The third main difference between the {classical} disturbing function and that of nearly polar orbits is the dependence on the parity of the resonance order and the corresponding universal binarity of the force amplitudes of resonant terms. To lowest order in $e$ and $\cos I$, all resonances $p$:$q$ with an even $o_r=|p-q|$ have force amplitudes that are quadratic with respect to eccentricity and constant with respect to inclination whereas all resonances $p$:$q$ with an odd $o_r=|p-q|$ have force amplitudes that are linear with respect to eccentricity. As was mentioned in the previous section, this curious behaviour stems from the presence of the two independent angles $f+\omega$ and $\Omega-\lambda^\prime$ in the relative distance $\Delta$.  We suspect, but cannot prove, that the property is related to the fact that for a given resonance, one must have prograde as well as retrograde arguments in the same polar series unlike the {classical} disturbing functions of nearly coplanar  prograde or retrograde orbits.

We gain further insight into the structure of the disturbing function for nearly polar orbits by seeking the natural variables with which polar motion {can} be studied.   To do this we recall that instead of using the standard orbital elements, Poincar\'e devised canonical action-angle variables for studying the three-body problem  that  have the property of including two actions related to eccentricity and inclination that vanish when motion is exactly  circular, coplanar and prograde. The Poincar\'e canonical variables are given as the three pairs:
\begin{eqnarray}
\Lambda&=&mna^2, \ \ \ \lambda=M+\omega+\Omega, \nonumber\\
\Gamma &=& mna^2\left[1-(1-e^2)^{1/2}\right], \ \ \ \gamma=-\omega-\Omega,\\
Z&=&m na^2\left(1-e^2\right)^{1/2}(1-\cos I), \ \ \ z=-\Omega, \nonumber
\end{eqnarray}
where $m$ is the particle's mass, $n$ its mean motion and $M$ its mean longitude. The appropriate variables for polar motion must have the property  that the actions related to eccentricity and inclination vanish when motion is exactly polar and circular. The Poincar\'e action $\Gamma$ satisfies this condition but not $Z$. The latter {can} be replaced by  $Z^\star=\Lambda-\Gamma-Z=m na^2\left(1-e^2\right)^{1/2}\cos I$, the {normal} component of angular momentum. The remaining variables are obtained from the following generating function:
\begin{equation}
F=(\lambda + z)\Lambda^\star+ (\gamma-z) \Gamma^\star-z Z^\star.
\end{equation}
Using $Z=\partial_z F$, $z^\star=\partial_{Z^\star}F$ etc, we find:
\begin{eqnarray}
\Lambda^\star&=&\Lambda, \ \  \ \lambda^\star=\lambda-\Omega= M+\omega,\nonumber\\
\Gamma^\star&=& \Gamma, \ \ \ \gamma^\star=-\omega,\label{polarpoincare}\\
Z^\star&=& m na^2\left(1-e^2\right)^{1/2}\,\cos I, \ z^\star=\Omega.\nonumber
\end{eqnarray}
It  can be seen that the choice of the correct variable $Z^\star$ modifies the mean longitude $\lambda^\star$, the longitude of pericenter $\gamma^\star$ and the angle $z^\star$ associated with the longitude of ascending node showing that the argument of pericentre is one of the natural angles that should describe polar motion.  For comparison, when we modified the Poincar\'e canonical variables in \cite{NamouniMorais15} to study retrograde resonances by choosing the new inclination action as  $Z^r=2(\Lambda-\Gamma)-Z$ so that $Z^r$ vanishes for exactly  coplanar retrograde motion,  the mean longitude and longitude of pericentre were modified to $\lambda^r=M+\omega-\Omega$ and $\gamma^r=\Omega-\omega$ thus producing  the natural angles with which retrograde resonances {can} be studied.  We note that our choice of the third canonical action is not unique. For instance instead of $m na^2\left(1-e^2\right)^{1/2}\,\cos I,$ one could employ $m na^2\left(1-e^2\right)^{1/2}\,(1-\sin I)$ which has the added advantage of being positive regardless of inclination. The corresponding new angles, however, are no longer function of the old angles, they will  also depend on $\Lambda^\star$, $\Gamma^\star$ and $Z^\star$. 

Using the new polar canonical angles, the argument of the cosine terms in the disturbing function (\ref{RY}) is transformed as $\phi^{p:q}_k=\phi-k\omega= p\lambda^\star-q(\lambda^\prime-z^\star)+k\gamma^\star$ implying a simple physical meaning that polar mean longitude need only be measured as if motion were two-dimensional. {The} particle's longitude of ascending node must be used {as} a reference line to measure the mean longitude of the planet as the latter two angles lie in the same plane. The remaining term $k\gamma^\star$ gives the $k$th-harmonic that {could be} excited by the planet.  Lastly, we also mention that the new polar canonical variables are related to the {classical} Delauney variables given as ($L=\Lambda^\star,$ $l=\lambda^\star+\gamma^\star$), ($G=\Lambda^\star-\Gamma^\star$, $g=-\gamma^\star$) and ($H=Z^\star$, $h=z^\star$). These variables were used by \cite{Kozai62} to study the secular evolution at large eccentricity and inclination in the three-body problem that will be discussed in the next section.

\section{Secular potential and validity domain of the disturbing function}\label{sec5}
The validity domain of the disturbing function of nearly polar orbits is related {to} the secular potential  that governs the long term dynamics of the particle. The reason is the large inclination of the particle's orbital plane relative to the planet's that {could} lead to large eccentricity and inclination oscillations. In the three-body problem with a planet on a circular orbit, secular evolution of a particle with a non-resonant orbit  is given by the Kozai-Lidov potential \citep{Kozai62,Lidov62}. Its integral expression \citep{Quinn90} {is} written with our notations as:
\begin{eqnarray}
R_{KL}&=&\frac{Gm^\prime}{\pi a^\prime \alpha^2(1-e^2)^{1/2}}\int_0^{2\pi}R^{-1/2}{k \, K(k)} r^2 df, \label{RKL}\\
k^2&=& \frac{4R}{(R+1)^2+z^2},\nonumber\\
R^2&=&r^2-z^2,\nonumber\\
z&=&r \sin I \sin(f+\omega),\nonumber
\end{eqnarray}
where $K$ is the complete elliptic integral of the first kind and $r=\alpha (1-e^2)/ (1+e \cos f)$ is the particle's orbital radius defined in Section \ref{sec2}. The Kozai-Lidov potential is the doubly averaged gravitational potential with respect to the particle's and planet's mean longitudes and does not involve any expansion with respect to eccentricity and inclination. As it depends on the sole angle $\omega$, one {can} use the Delauney variables and find that both the semimajor axis $\alpha$ and the {normal} component of angular momentum $\left(1-e^2\right)^{1/2}\,\cos I$ are constants of secular evolution. Motion generated by the Kozai-Lidov potential thus occurs in the eccentricity-argument of pericentre plane. In writing the expression  (\ref{RKL}) no assumption was made on the inclination; therefore it {can} equally be $\leq$ or $\geq 90^\circ$. Close examination of (\ref{RKL}) and {noting the manner in which} the {normal} coordinate, $z$,  enters its expression reveal that the secular structures in the $e\omega$--plane  for prograde and retrograde orbits are identical and one can study the former and deduce the latter because of the potential's reflection symmetry with respect to the polar plane.

We study the validity domain  of the disturbing function by comparing the secular potential it produces with the Kozai-Lidov potential as in essence the former is an expansion of order $N$ of the latter with respect to eccentricity and inclination cosine for nearly polar orbits. This comparison is valid only in the absence of  resonance libration but should illustrate the typical values of eccentricity and inclination cosine where the fourth order series {can} be used. The literal expansion of Section \ref{sec2} shows that the secular potential to order $N$ in eccentricity and inclination cosine is given as:
\begin{equation}
\bar{R}_s=\frac{1}{2}b_\frac{1}{2}^{00}(\alpha)+\!\!\!\!\!\!\!\sum_{
{\scriptsize\begin{array}{c}
0\leq k,n\leq N \\
k\leq m\leq N\\
k,m, n  \ {\rm even}\\
m+n=N
\end{array}}} \!\!\!\!s^{k}_{mn}(\alpha) \, e^{m} \cos^{n} I\ \cos (k\omega),\label{SecPot}
\end{equation}
where the corresponding force coefficients $s^{k}_{mn}(\alpha)$ are given in Table \ref{tsec} for $N=4$. 

We assess the possible large variations of eccentricity and inclination produced  by the secular potentials for nearly polar orbits by plotting in the $e\omega$--plane the level curves of $R_{KL}a^\prime/Gm^\prime$ and $\bar{R}_s$ for two initially circular orbits located at $\alpha=2$  and $\alpha=0.5$ so as to illustrate the effects of internal and  external perturbers respectively. Since these locations are near the 1:3 and 3:1 resonances respectively, both secular potentials reflect the dynamics of  circulating orbits to first order in the perturber's mass.\footnote{Near-mean motion resonances introduce an additional secular potential effect for circulating orbits whose amplitude is of second order in the perturber's mass \citep{Hagihara}.} We will see in Section \ref{sec6} the evolution of eccentricity and inclination of resonant orbits is somewhat modified by the critical arguments' libration.

\begin{figure*}
\begin{center}
{ 
\includegraphics[width=40mm]{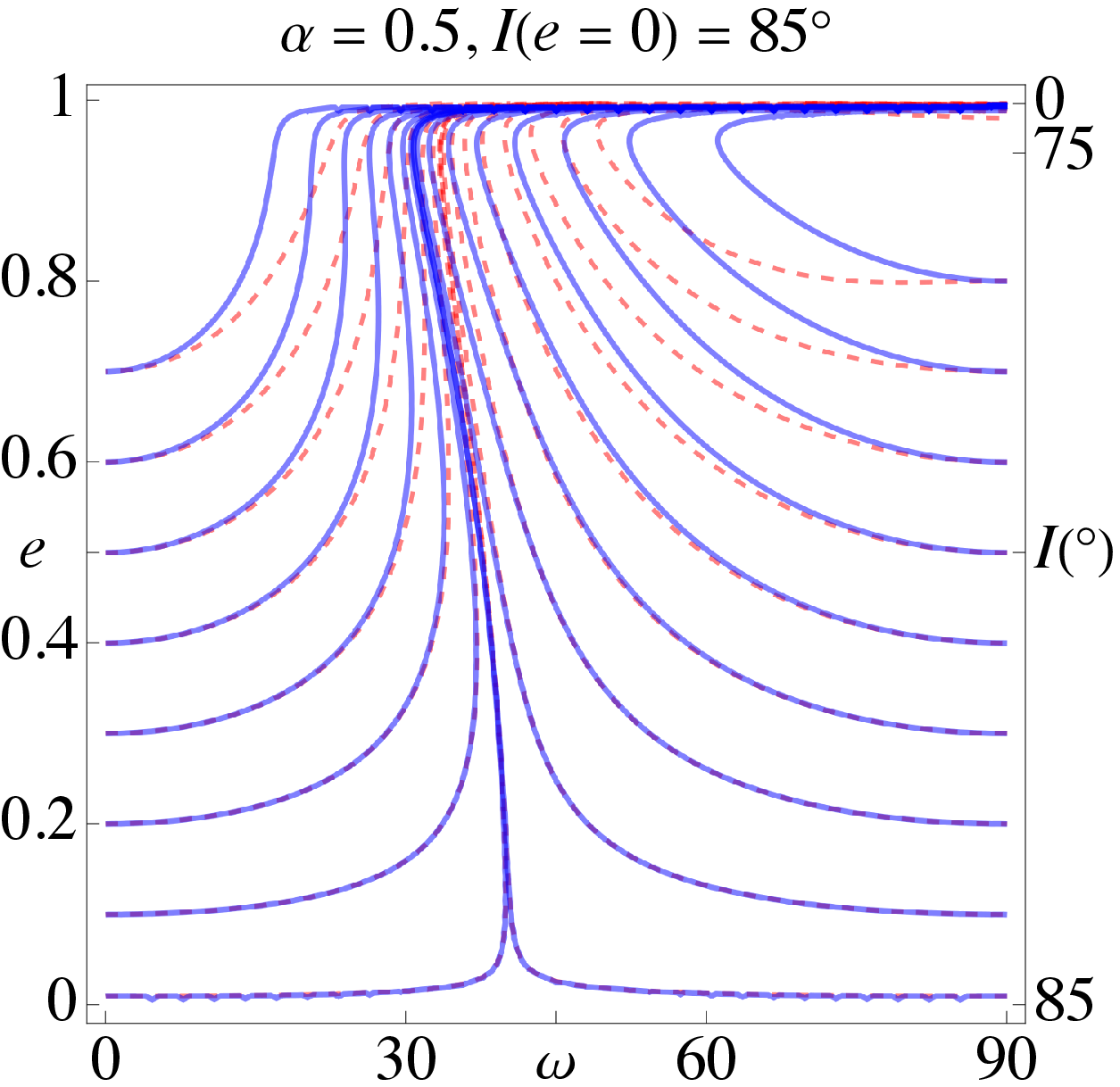}\ \ \includegraphics[width=40mm]{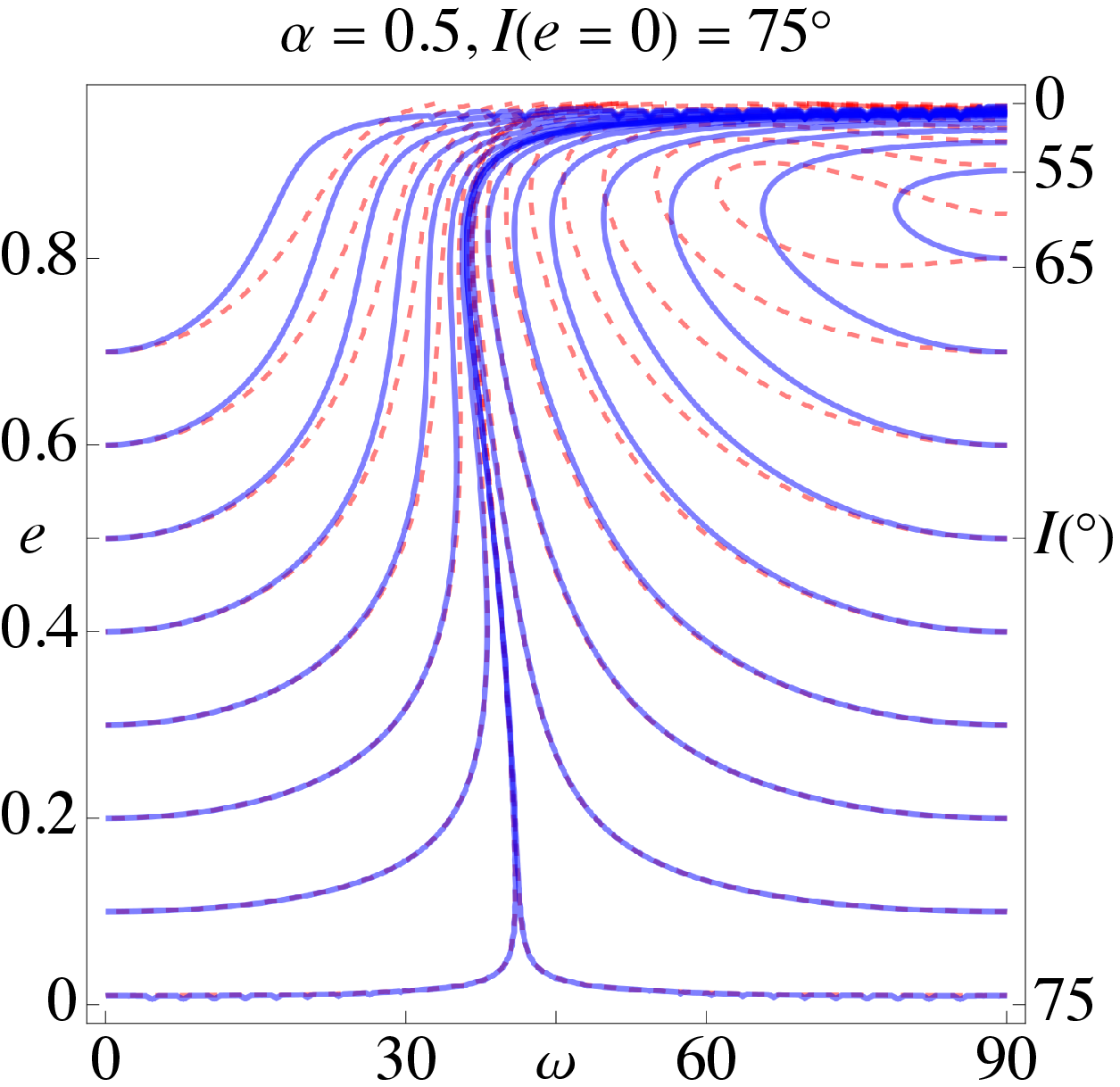}\ \ \includegraphics[width=40mm]{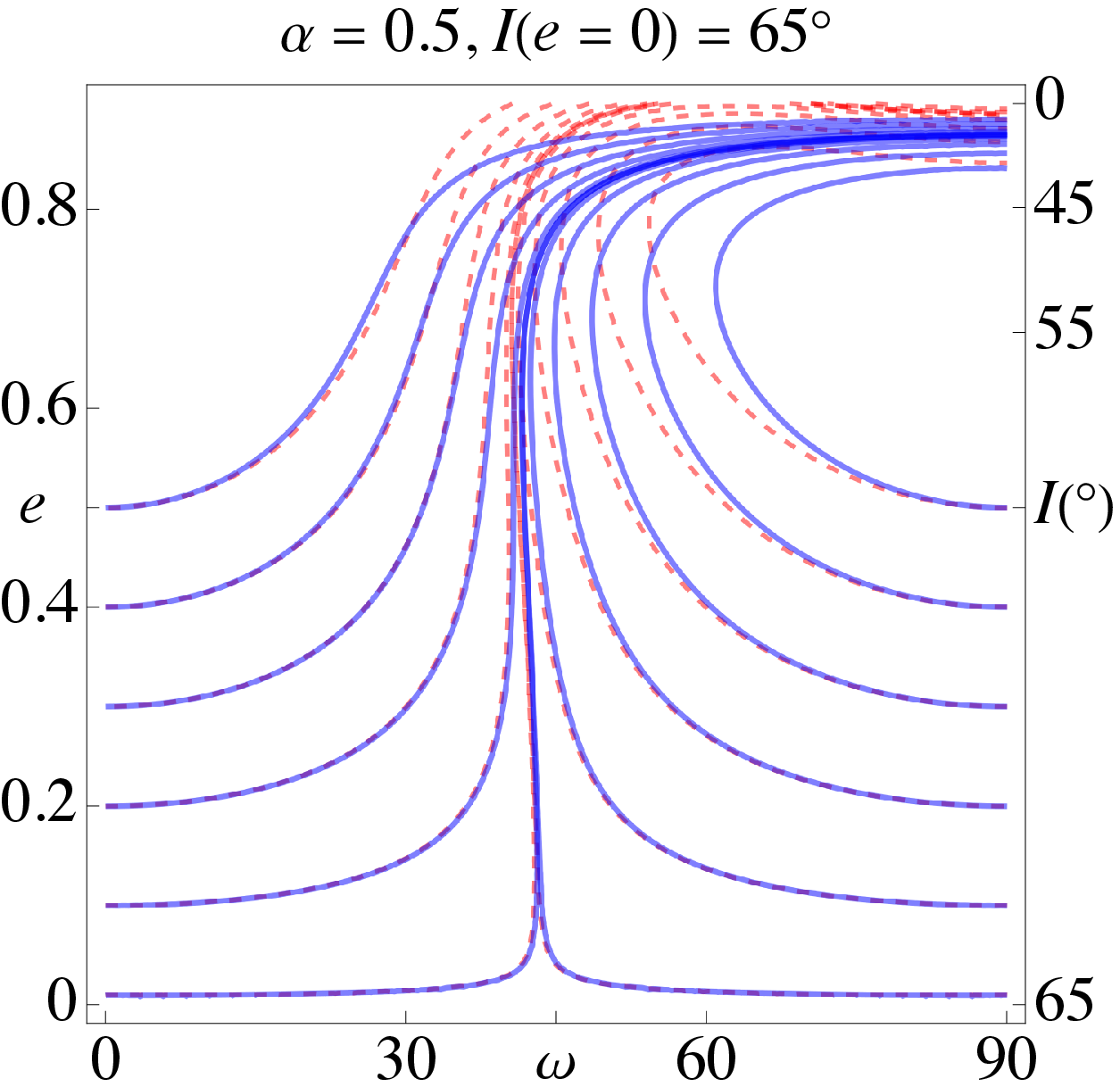}\ \ \includegraphics[width=40mm]{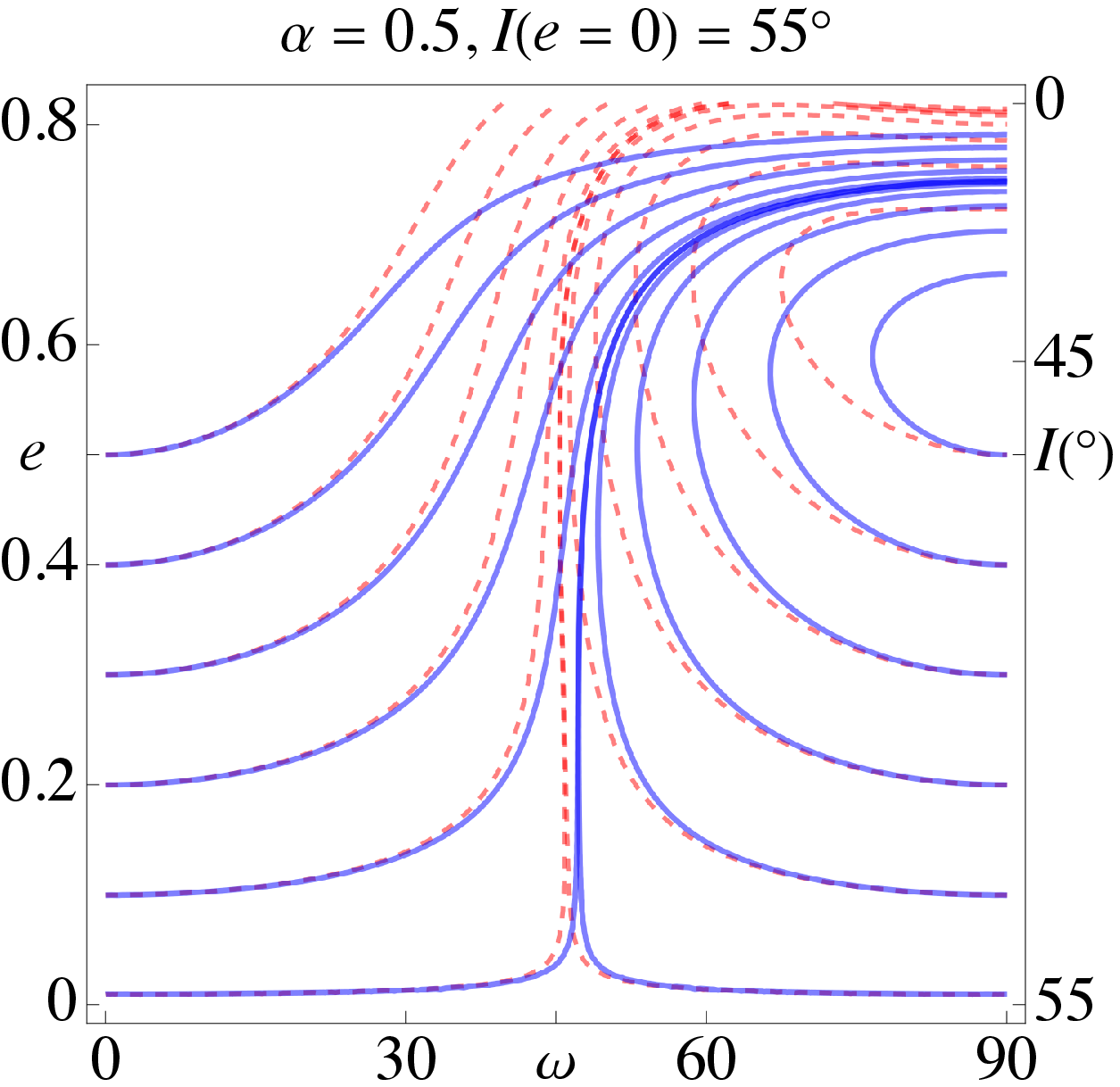}\\[2mm]
\includegraphics[width=40mm]{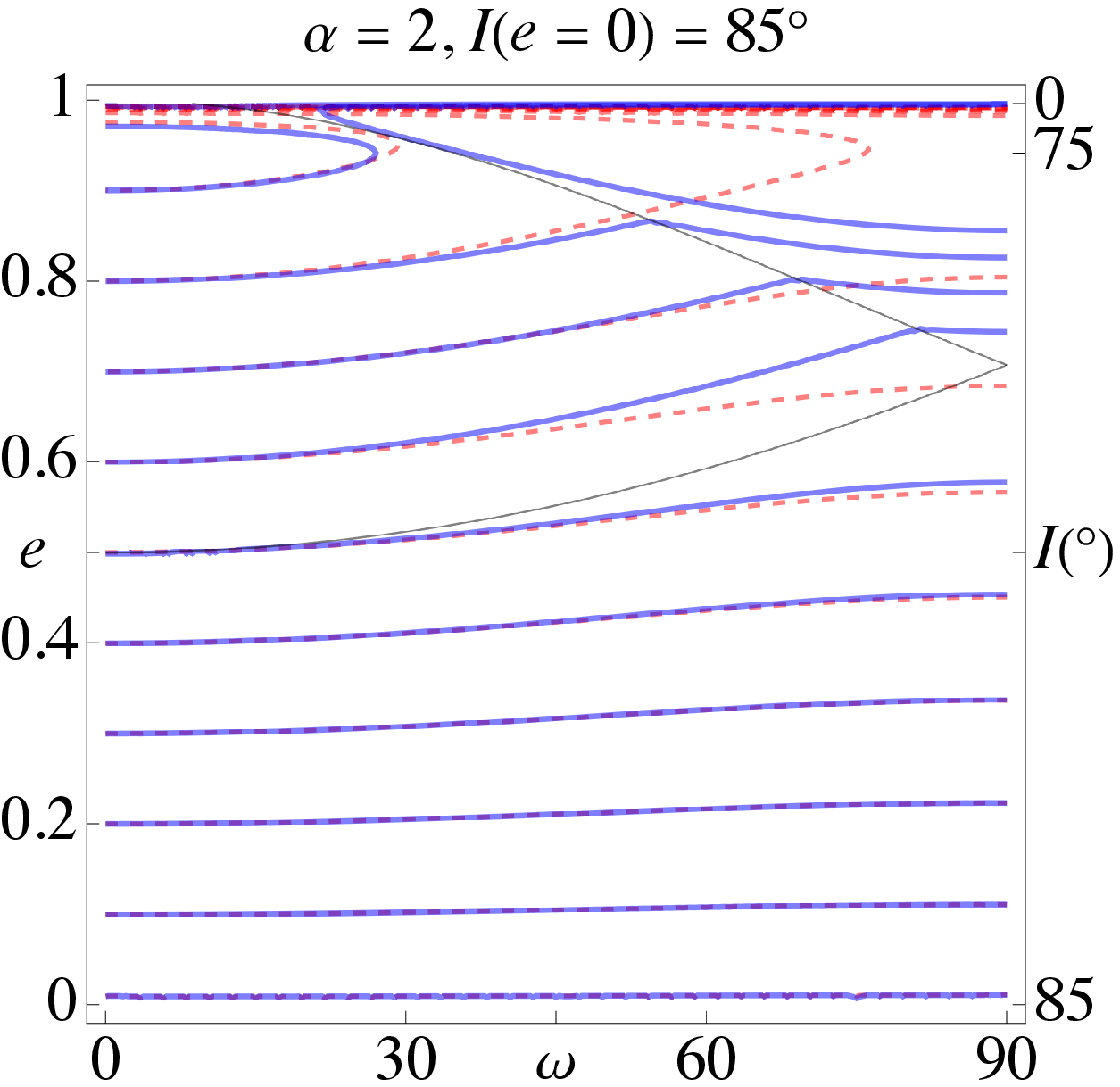}\ \ \includegraphics[width=40mm]{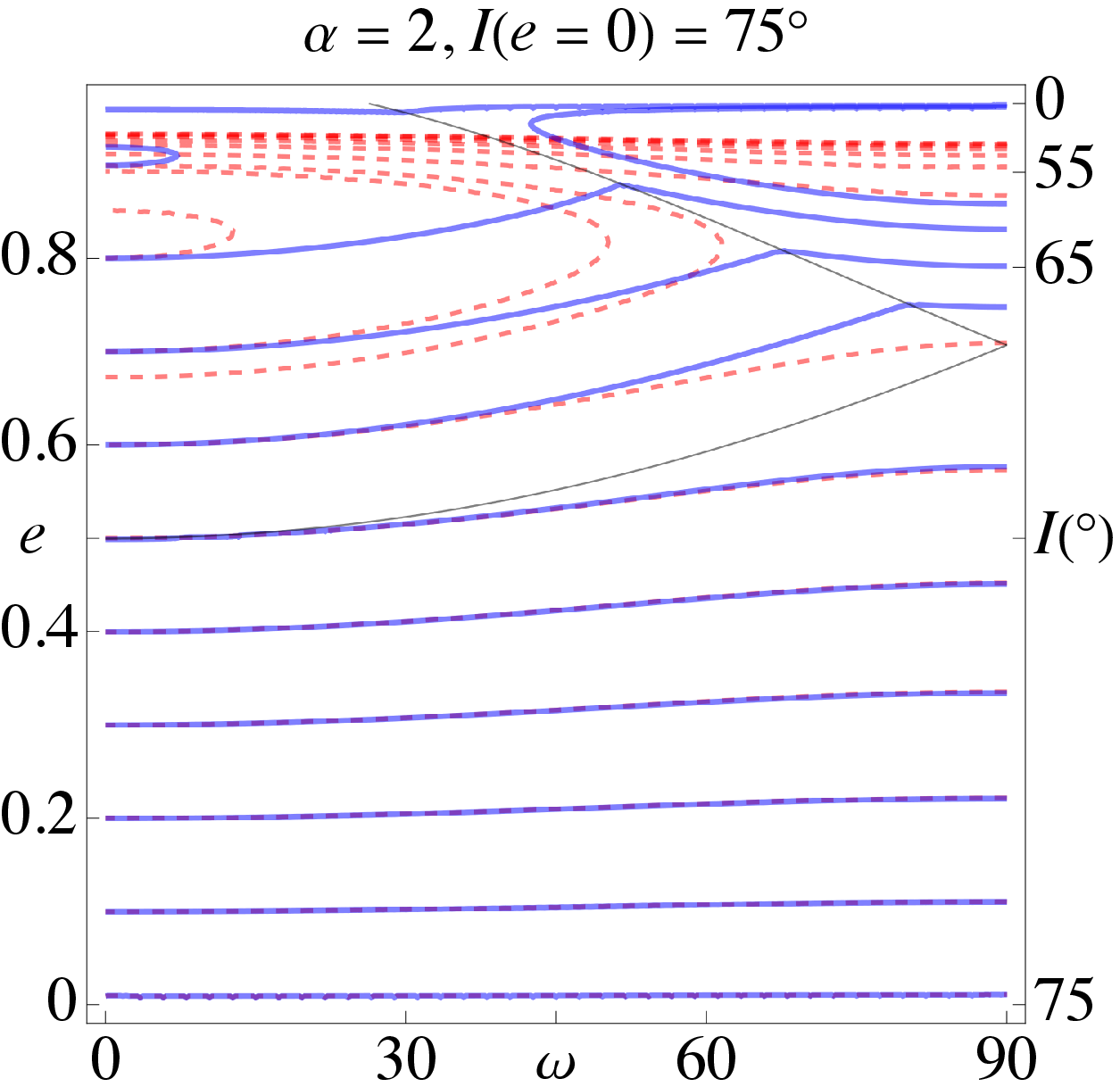}\ \ \includegraphics[width=40mm]{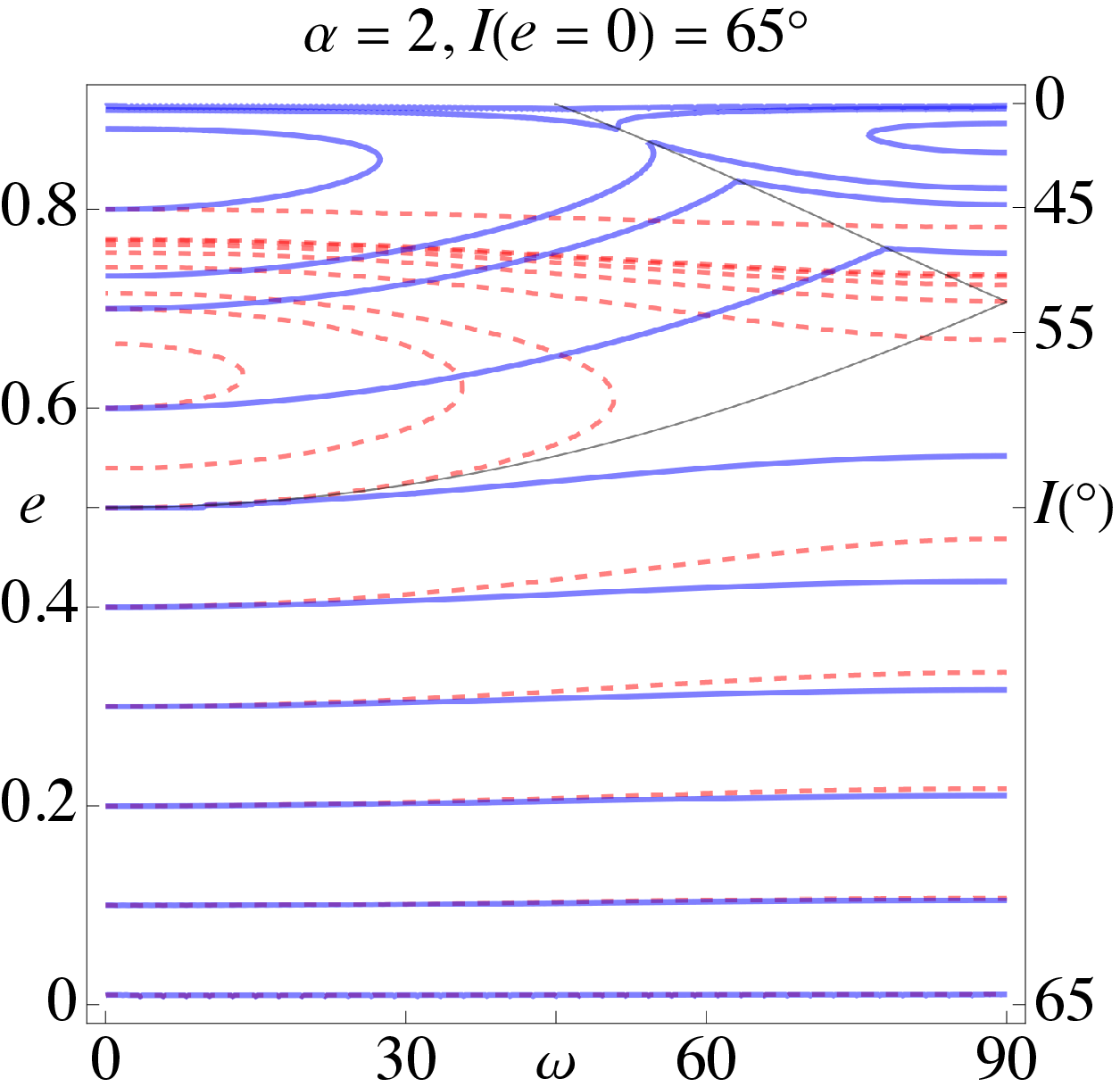}\ \ \includegraphics[width=40mm]{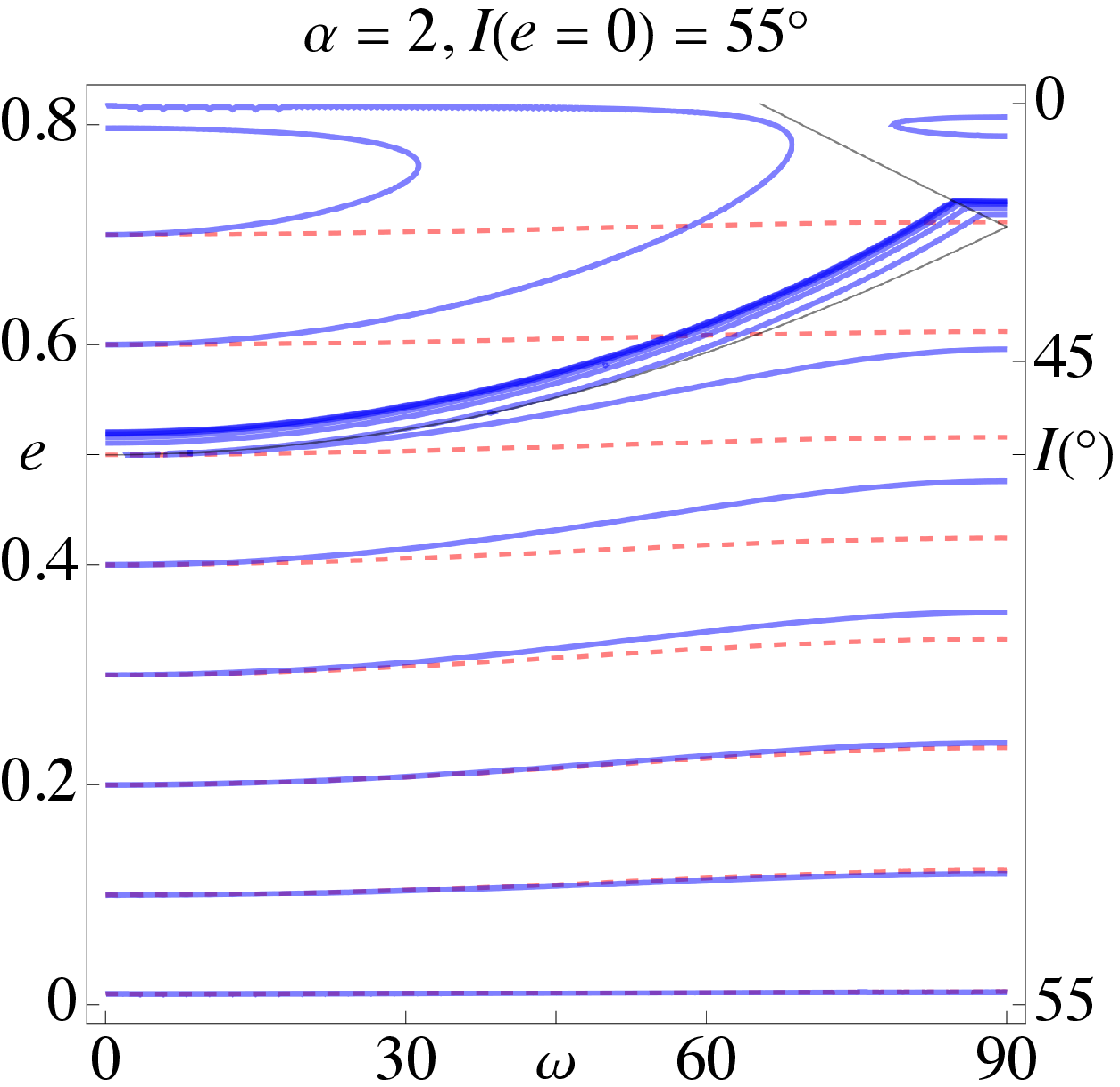}\\
}
\caption{Level curves of the secular potentials for an initially circular orbit with semimajor axis ratio $\alpha=0.5$ (top panels) and $\alpha=2$ (bottom panels) for four inclination values $I=85^\circ, 75^\circ, 65^\circ, 55^\circ$. The solid blue lines represent the Kozai-Lidov potential (\ref{RKL}) and the dashed red lines the 4th order polar secular potential (\ref{SecPot}). The thin lines represent the collision singularity.}\label{f1}
\end{center}
\end{figure*}

The initial inclinations are taken as $I(e=0)=85^\circ$, $75^\circ$, $65^\circ$ and $55^\circ$ (Figure \ref{f1}). Owing to the symmetry with respect to the polar plane, these values produce {exactly the} same $e\omega$--portraits as $I(e=0)=95^\circ$, $105^\circ$, $115^\circ$ and $125^\circ$ respectively. The only difference is the inclination range that reads  [$I(e=0)$:$180^\circ$] for retrograde orbits instead of [$I(e=0)$:$0^\circ$] for prograde orbits. 
The various level curves in each panel correspond to additional orbits with a {normal} angular momentum $Z^\star$ equal to that of the reference particle namely $\left(1-e^2\right)^{1/2}\,\cos I=\cos[ I(e=0)]$.   
It is seen on the full Kozai-Lidov potential that particles located inside the planet's orbit ($\alpha=0.5$) are unstable in the sense that they inevitably reach a near-unit eccentricity corresponding to an orbit that is nearly coplanar with the planet. In the absence of mean motion resonance libration, the time it takes {for} a particle on a nearly polar orbit with a moderate  eccentricity to reach a nearly coplanar orbit is {long} (see the example in Section \ref{sec6}). The fourth order secular potential is found to reproduce the dynamics quite well for $e\leq 0.5$ and $I(e=0)\geq65^\circ$ in the case of an external perturber (i.e. $\alpha=0.5$). The potential $\bar{R}_s$ {can} be used to follow the dynamics on timescales {shorter} than the libration around the Kozai-Lidov resonance at $\omega=90^\circ$. For timescales comparable to the libration time, one needs to push the expansion order to larger values so as  to improve the dynamics' rendition. Particles outside the planet's orbit (i.e. $\alpha=2$) fall into two types of motion. Up to an eccentricity $e=0.5$, the argument of pericentre circulates and eccentricity and therefore also inclination have  small amplitude variations. The validity domain of the fourth order polar disturbing function   is $e\leq 0.5$ for $I(e=0)\geq75^\circ$  and $e\leq 0.2$ for $55^\circ\leq I(e=0)\leq65^\circ$.  Above $e=0.5$, eccentricity can be made close to unity by the two Kozai-Lidov resonances at $\omega=0$ and $90^\circ$ (with the obvious exception of the resonance centers vicinity) and the use of the disturbing function would require a larger (than 4) expansion order like {in} the case of the external perturber.

In order to understand the secular evolution of the line of nodes that was absent from the previous analysis, we use the secular potential  (\ref{SecPot}) and apply it to the motion of  a massless particle perturbed by an internal planet (i.e. $\alpha>1$) when the {particle}'s orbit is far from the Kozai-Lidov resonances.  The corresponding $e\omega$--curve is therefore located in the bottom part of the second row plots of Figure \ref{f1}.  The secular potential  (\ref{SecPot}) restricted to second order in eccentricity and inclination cosine will suffice to follow the motion of the longitude of ascending node. It is written as:
\begin{equation}
\bar{R}_s= s_{20}^0e^2 + s^2_{20}e^2\cos 2\omega+ s^0_{02} \cos^2I.  \label{2orderPotSec}
\end{equation}
where we removed the first term of (\ref{SecPot}) as it does not influence eccentricity and inclination. The Lagrange planetary equations {can} be written in terms of $e$, $\omega$, $I$ and $\Omega$ (\cite{BC61} page 289) and truncated for nearly polar orbits to  lowest order in eccentricity and inclination cosine to give:
\begin{eqnarray}
\dot e = -(na^2e)^{-1}\partial_\omega R_s, \ \ \ \dot \omega = (na^2e)^{-1}\partial_e R_s, && \label{Lap-eo}\\
\dot I = -(na^2)^{-1}\partial_\Omega R_s, \ \ \ \dot \Omega = (na^2)^{-1}\partial_I R_s, \ \ \ && \label{Lap-IO} 
\end{eqnarray}
where $n=(GM_\star a^{-3})^{1/2}$ is the particle's mean motion and $R_s=Gm^\prime a^{\prime -1}\bar{R}_s$ the fully dimensional secular potential.  We note how the $I\Omega$--equations (\ref{Lap-IO}) differ form those of nearly coplanar orbits in that the variation rates are not inversely proportional to inclination unlike the $e\omega$--equations (\ref{Lap-eo}) that keep their {classical} form. Furthermore, by using the potential (\ref{2orderPotSec}) in the Lagrange equations, it is found that the inclination $I$ is a constant of motion implying  far smaller variations for $I$  than for $e$ when  resonant terms are included. The longitude of ascending node's  variation rate  for nearly polar orbits is {also constant as it depends only on $I$} and given as:
\begin{equation}
\dot \Omega=-\frac{ n \alpha m^\prime s^0_{02}\sin 2I}{M_\star}.
\end{equation}
The  secular force coefficient that enters the expression of $\dot\Omega$ {can} be approximated by $s^0_{02}\sim 11(\alpha-1)(6.5+370(\alpha-1)^{2.1})^{-1}$ for $1<\alpha\leq 5$. Therefore as $s^0_{02}>0$,  the line of nodes of prograde nearly polar orbits regresses whereas that of retrograde nearly polar  orbits precesses. The nearer to polar motion the smaller the variation rate of the longitude of ascending node. These results are confirmed in the next section.

\section{Examples of polar resonance}\label{sec6}
In this section, we provide an illustration of how the disturbing function helps us to identify the correct resonant arguments associated with the polar motion of a particle that interacts with a Neptune-mass planet on a circular-orbit ($m^\prime/M_\star=  5.12\times 10^{-5}$). We will not develop a comparison of the analytical polar disturbing function using the Lagrange equations  with numerical integrations  as it is beyond the scope of this work.  Instead, we integrate the full equations of motion only to follow the evolution of the particle's orbit and show a variety of polar resonances. We shall consider the following  resonances   1:3, 3:1, 2:9,  and 7:9. 

\begin{figure*}
\begin{center}
{ 
\includegraphics[width=170mm]{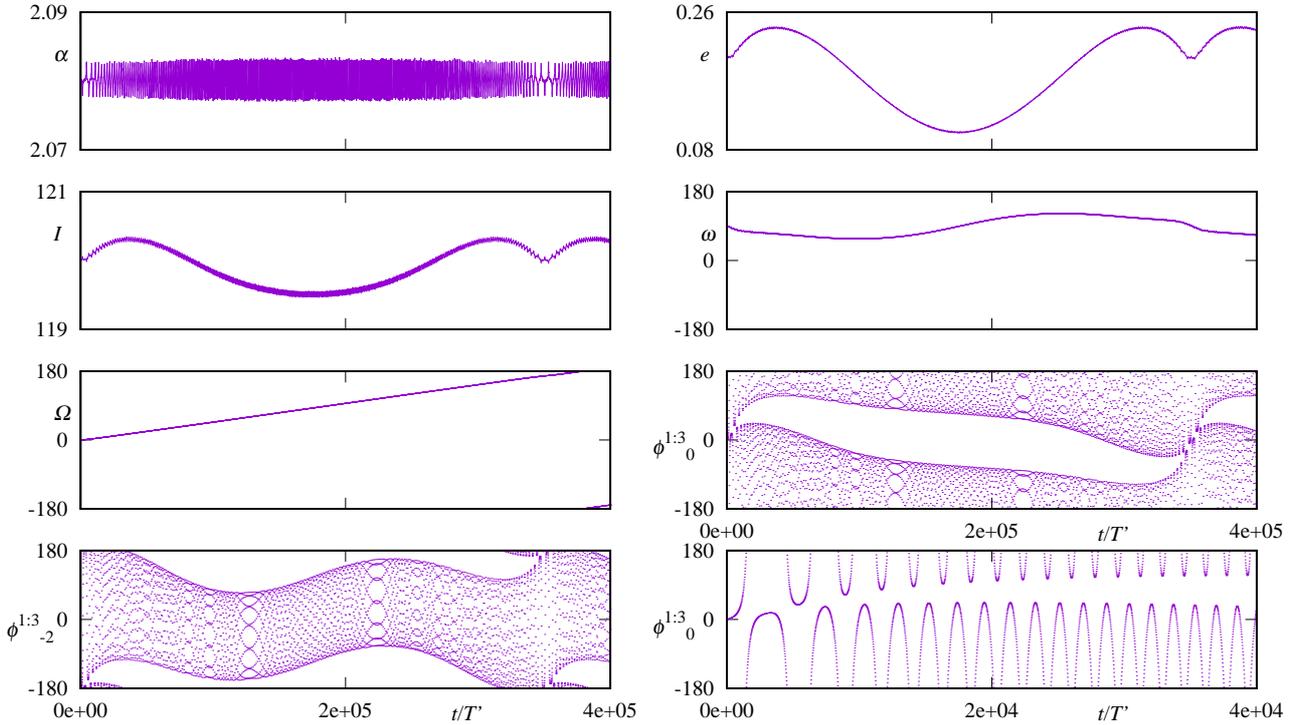}
\caption{Orbital elements $\alpha$, $e$, $\omega$, $I$, $\Omega$ and resonant arguments  $\phi^{1:3}_0$, $\phi^{1:3}_{-2}$ as a function of time for  a particle at the outer 1:3 resonance with a Neptune mass planet. Initial parameters are:  eccentricity $e=0.2$, inclination $I=120^\circ$, longitude of ascending node $\Omega=0^\circ$, argument of pericentre $\omega=90^\circ$ and relative mean longitude $\lambda-\lambda^\prime= 0^\circ$. The bottom right panel is a zoom of $\phi^{1:3}_0$ for $0\leq t/T^\prime\leq 10^4$. }\label{f2}
}
\end{center}
\end{figure*}

We learned in Section \ref{sec3} that the arguments that enter the disturbing function are of the form $\phi^{p:q}_k=\phi-k\omega$ where $\phi=q \lambda -p \lambda^\prime +(p-q) \Omega$. The fundamental mode $k=0$ occurs only for even-order resonances. It is a pure-inclination term that in principle could librate regardless of eccentricity as its first order force amplitude is  $c^0_{00}+c^0_{01}\cos I$ giving the resonance a pendulum-like dynamical structure almost independent of inclination for polar-like orbits $I\sim 90^\circ$.  However nearly polar orbits have a large relative inclination with respect to the planet's orbit and that in turn {can} force a coupling of eccentricity and inclination variations similar to what was seen in the previous Section with the Kozai-Lidov resonance. We therefore illustrate the fundamental mode $k=0$ by placing the particle directly in the Kozai-Lidov resonance that is coupled to the outer 1:3 mean motion resonance thus ensuring that the argument of perihelion $\omega$ is stationary and allowing the $k=0$ mode to librate. Figure \ref{f2} shows as functions of time, normalized to the planet's orbital period $T^\prime$, the evolution of the orbital elements along with the resonant angles $\phi^{1:3}_k=3 \lambda-\lambda^\prime-2\Omega-k\omega$ for $k=0$ and $-2$.  The particle's initial orbital elements are $e=0.2$, $\omega=90^\circ$, $I=120^\circ$, $\Omega=0^\circ$, and $\alpha=3^{2/3}$ is the nominal resonance location. The argument $\phi^{1:3}_{0}$  librates with a variable  period starting from a maximum of $4000\,T^\prime$  and evolving to a minimum of $1000\,T^\prime$. A similar behaviour is forced on the $\phi^{1:3}_{-2}$ argument because of secular resonance.  The 1:3 resonance also modifies the $e\omega$--secular structure in that it allows the Kozai-Lidov resonance at $\omega=90^\circ$ to occur at a much lower eccentricity than that of non-resonant orbits considered in Section \ref{sec5}. 
 However, whereas the eccentricity's variations are moderate, the inclination's are quite small and the line of nodes precesses linearly as the orbit is retrograde confirming the results of Section \ref{sec5}. 

\begin{figure*}
\begin{center}
{ 
\includegraphics[width=170mm]{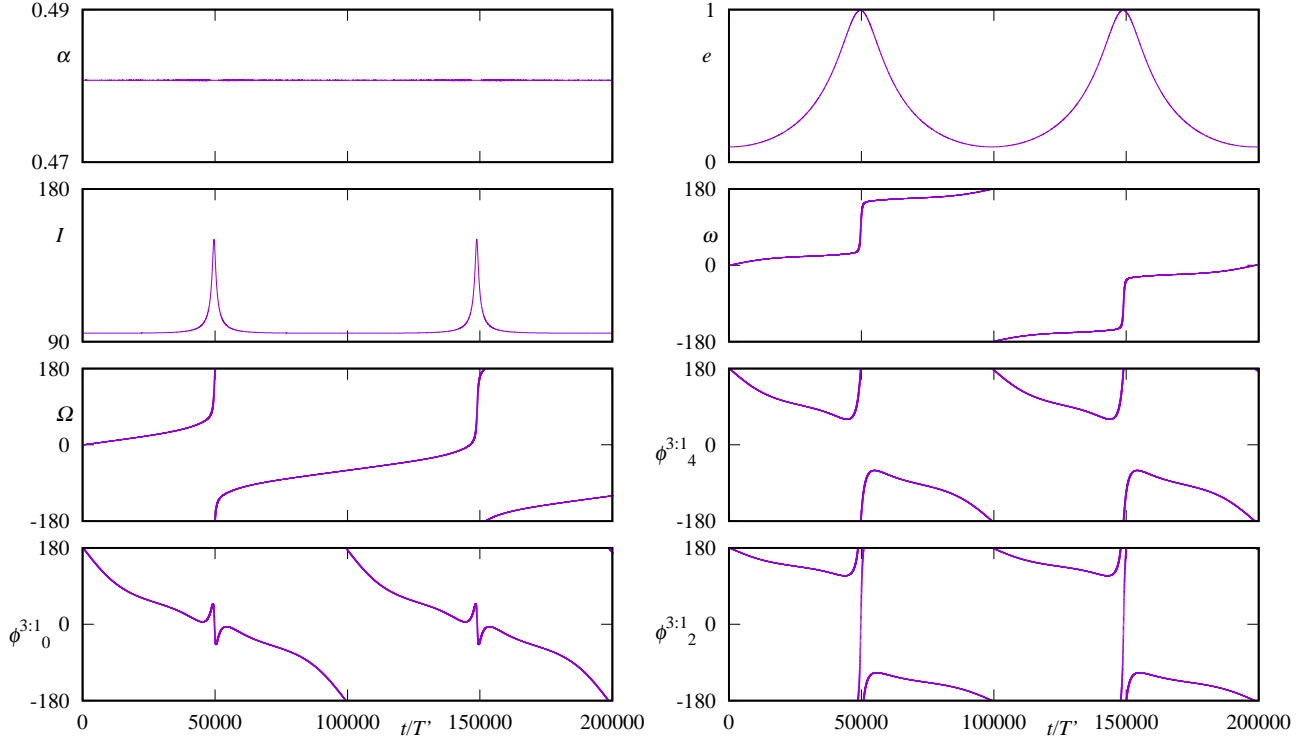}
\caption{ Orbital elements $\alpha$, $e$, $\omega$, $I$, $\Omega$ and resonant arguments  $\phi^{3:1}_0$, $\phi^{3:1}_{2}$ and  $\phi^{3:1}_{4}$ as a function of time for a particle at the inner 3:1 resonance with a Neptune mass planet. Initial parameters are:  eccentricity $e=0.1$, inclination $I=95^\circ$, longitude of ascending node $\Omega=0^\circ$, argument of pericentre $\omega=0^\circ$ and relative mean longitude $\lambda-\lambda^\prime= 180^\circ$. }\label{f3}
}
\end{center}
\end{figure*}

On the subject of the effect of the Kozai-Lidov secular potential on polar orbits, we also examine the evolution of a particle that librates in the inner 3:1 resonance located at  $\alpha=3^{-2/3}$ to give an example with an external perturber and ascertain the similarities and differences with the secular evolution of the non-resonant orbits discussed in Section \ref{sec5}. The possible resonant critical arguments now read $\phi^{3:1}_k=\lambda-3\lambda^\prime+2\Omega-k\omega$ where $k$ is an even integer. Placing a particle at the bottom of the $e\omega$-plane of the first top panel of Figure \ref{f1} with $e=0.1$,  $\omega=0$, $\Omega=0$ and choosing an inclination $I=95^\circ$ whose secular structure away from resonance is identical to that of $I=85^\circ$ (as explained at the beginning of Section \ref{sec5}), we show in Figure \ref{f3} the evolution of orbital elements as a function of time.   As expected from the effects of the secular potential of an external perturber, {the  argument of pericentre circulates periodically in the narrow {strip} adjacent to the Kozai-Lidov resonances where the eccentricity reaches a maximum value $e_{m}=0.995$.}  However the minimum inclination that should have been $180^\circ$  at maximum eccentricity if the particle were circulating instead of librating in mean motion resonance is now reduced to $150^\circ$  indicating that the conserved quantity involving inclination is no longer the {normal} component of angular momentum like that of the Kozai-Lidov potential of non-resonant orbits. 
In effect, of  the three critical  arguments $\phi^{3:1}_{0}$, $\phi^{3:1}_{2}$, $\phi^{3:1}_{4}$ the particle is found to stably librate in the $k=4$ resonance with a $120^\circ $-amplitude and  $10^5\,T^\prime$-period explaining why its secular evolution is not completely described by the Kozai-Lidov potential. { It is also interesting to note how other arguments display quasi-librations. For instance, $\phi^{3:1}_{2}$ follows the libration of $\phi^{3:1}_{4}$ over half the libration period only {to} circulate rapidly at maximum eccentricity. In a complementary way, the  argument $\phi^{3:1}_{0}$ circulates during  $\phi^{3:1}_{4}$'s libration and briefly librates at maximum eccentricity. }

With the next example, we illustrate how a resonance of order $o_r \gg 1$ {can} display librations with $|k|<o_r$ and consequently a force amplitude $\propto e^{|k|}$, a property {inherent} to the polar disturbing function first encountered in Section \ref{sec3} and discussed in Section \ref{sec4}.  We therefore {examine} the outer 2:9 resonance example (\ref{2:9}) located at $\alpha=(9/2)^{2/3}$  that has an odd resonance order, $o_r=7$, and possible resonant arguments  $\phi^{2:9}_{k}= 9\lambda-2\lambda^\prime-7\Omega-k\omega$, where $|k|\geq1$ is an odd integer, and force amplitudes $c^k_{|k|0}(2,9)\, e^{|k|}$ to lowest order in eccentricity. 

\begin{figure*}
\begin{center}
{ 
\includegraphics[width=170mm]{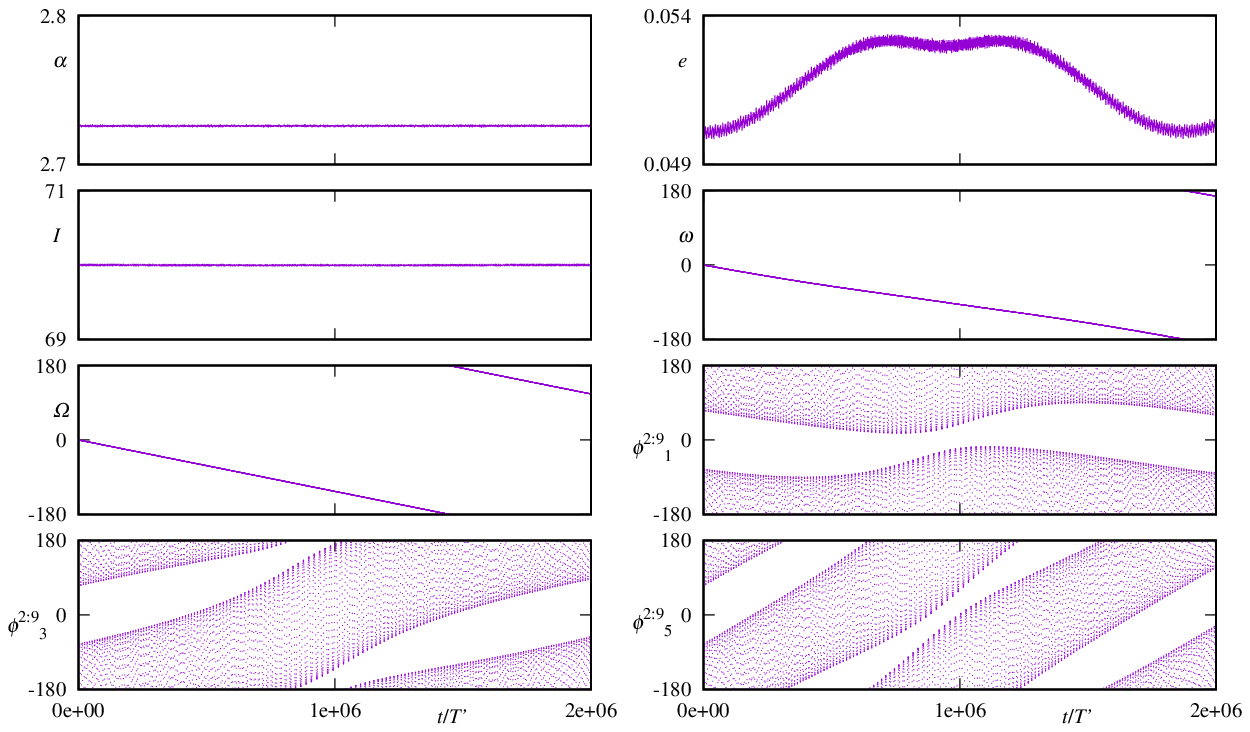}
}
\caption{Orbital {elements'} and resonant angles' evolution of particle at the outer 2:9 resonance with a {Neptune-mass} planet. Libration occurs for $\phi^{2:9}_1$. Initial parameters are:  eccentricity $e=0.05$, inclination $I=70^\circ$, longitude of ascending node $\Omega=0^\circ$, argument of pericentre $\omega=0^\circ$ and relative mean longitude $\lambda-\lambda^\prime= 90^\circ$.}\label{f4}
\end{center}
\end{figure*}

\begin{figure*}
\begin{center}
{ 
\includegraphics[width=170mm]{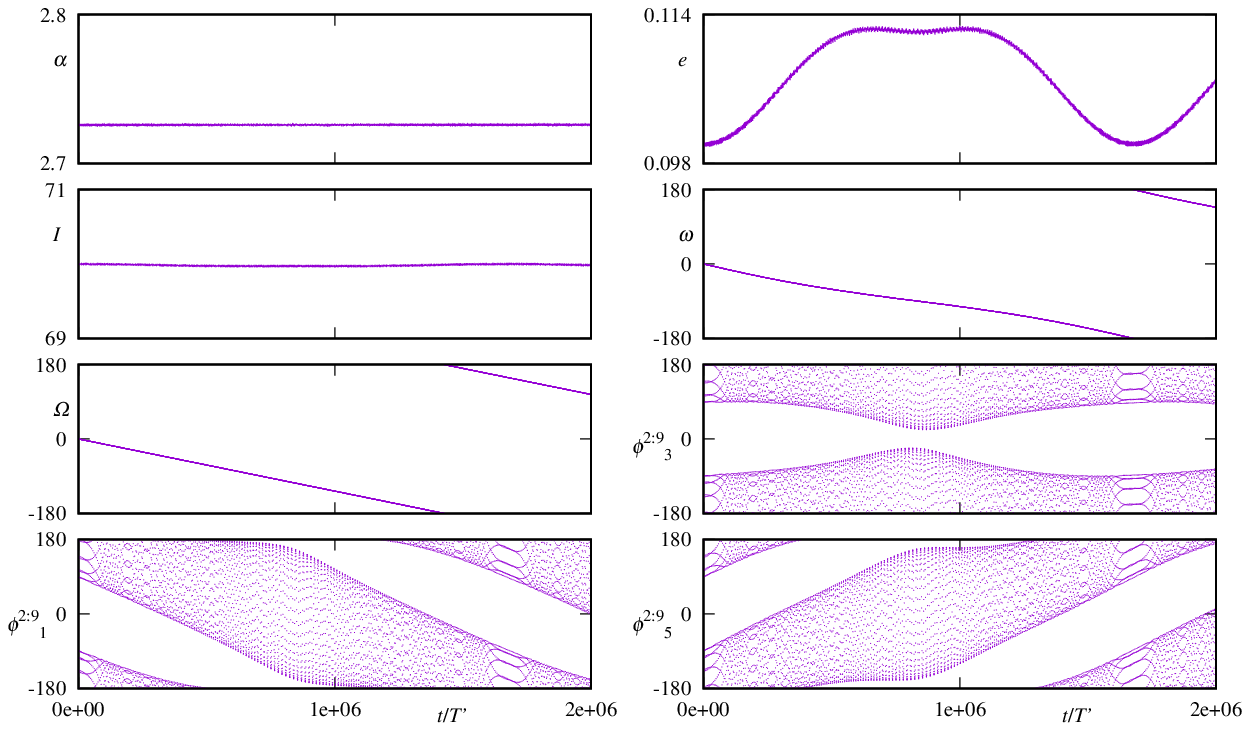}
}
\caption{Orbital {elements'} and resonant angles' evolution of particle at the outer 2:9 resonance with a {Neptune-mass} planet. Libration occurs for $\phi^{2:9}_3$. Initial parameters are:  eccentricity $e=0.1$, inclination $I=70^\circ$, longitude of ascending node $\Omega=0^\circ$, argument of pericentre $\omega=0^\circ$ and relative mean longitude $\lambda-\lambda^\prime= 90^\circ$.}\label{f5}
\end{center}
\end{figure*}

\begin{figure*}
\begin{center}
{ 
\includegraphics[width=170mm]{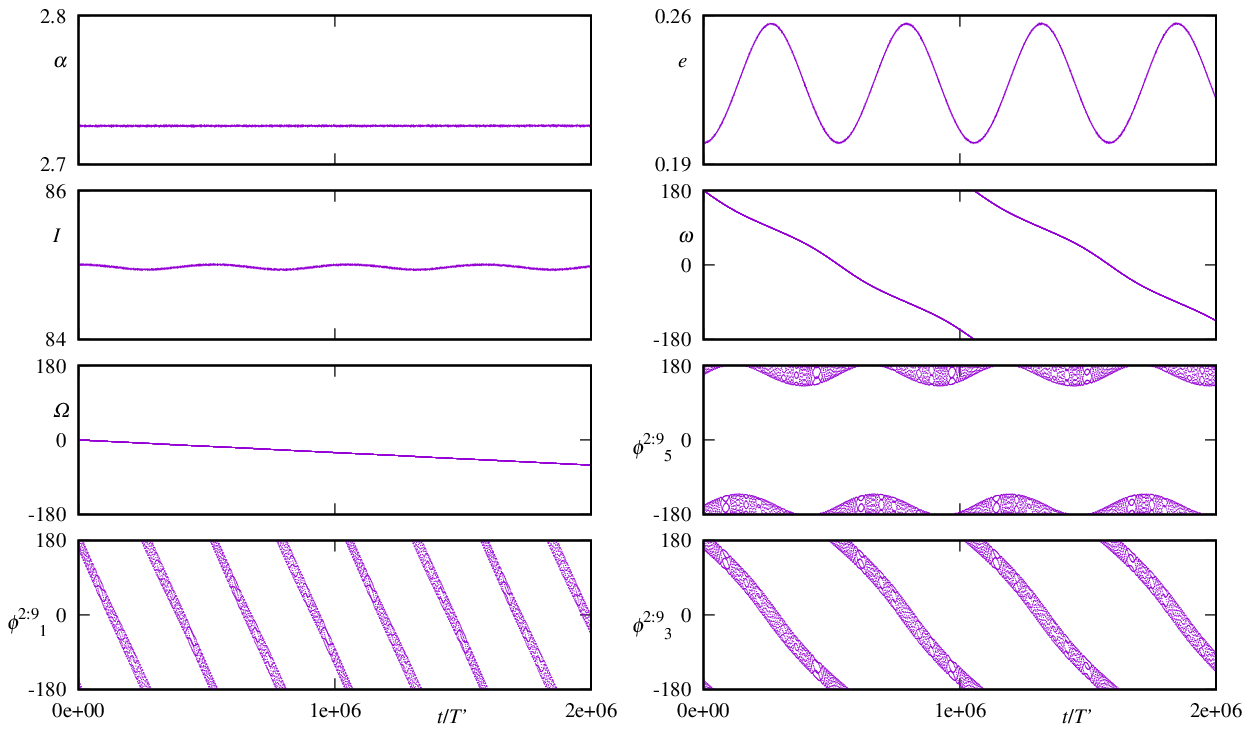}
}
\caption{Orbital {elements'} and resonant angles' evolution of particle at the outer 2:9 resonance with a {Neptune-mass} planet. Libration occurs for $\phi^{2:9}_5$. Initial parameters are:  eccentricity $e=0.2$, inclination $I=85^\circ$, longitude of ascending node $\Omega=0^\circ$, argument of pericentre $\omega=180^\circ$ and relative mean longitude $\lambda-\lambda^\prime= 0^\circ$.}\label{f6}
\end{center}
\end{figure*}

Figures \ref{f4}, \ref{f5} and \ref{f6} show the evolution of the orbital elements as well as the arguments $\phi^{2:9}_{k}$ for $k=1,\ 3,$ and 5 for three different initial conditions. With the initial parameters $I=70^\circ$, $e=0.05$, $\varpi=0^\circ$ and $\Omega=0^\circ$, the particle librates with the critical argument $\phi^{2:9}_1$ while the other arguments circulate (Figure \ref{f4}).   More precisely, the libration involves two periods: a short one $9800\,T^\prime$ and  a longer modulation $2\times 10^6 \,T^\prime$. The first period is  from the  fundamental mode $\phi^{2:9}_0$  that circulates on the second timescale. We know this because the fundamental mode is not related to eccentricity and {consequently is not influenced by its secular evolution.}   {The longer period corresponds to the  time necessary to counter the fundamental mode's long term slow circulation with the slow drift of the argument of pericentre $\omega$, giving rise to the resonant angle $\phi^{2:9}_{1}$. } {We note {that} the resonant arguments $\phi^{2:9}_3$ and $\phi^{2:9}_5$ display the fast libration of the fundamental mode's short {period}, yet each is circulating with a slow rate for $\phi^{2:9}_3$ and slightly faster for $\phi^{2:9}_5$. This shows that when identifying Centaurs and TNOs in polar resonance, one must integrate their orbits {over} long timespans so as not to {misinterpret} evolutions such as those described for  $\phi^{2:9}_3$ and $\phi^{2:9}_5$ as true librations in resonance.}

Increasing the particle's eccentricity to $e=0.1$ has two effects: the argument of perihelion's {regression} is faster (see below)  leading to resonance with the critical argument $\phi^{2:9}_3$ (Figure \ref{f5}). The two {libration} periods are decreased: the short one to $4500\,T^\prime$ and the long one to $1.66\times10^6\,T^\prime$. 
{The cautionary note that we pointed out in the previous example is still valid for the new  eccentricity as the arguments $\phi^{2:9}_1$ and $\phi^{2:9}_5$ circulate but display the librations associated with the short period. The situation is even more misleading for $\phi^{2:9}_5$ because if the integration timespan were restricted to the interval [7,10]$\times 10^5\, T^\prime$, the argument would seem to be genuinely librating.} 

For the example in Figure \ref{f6}, we increased the inclination {to } $I=85^\circ$ and eccentricity to $e=0.2${\bf,} resulting in the libration of the critical argument $\phi^{2:9}_5$ with a small amplitude and the two periods $1400\,T^\prime$ (fast one associated with the fundamental mode $k=0$) and  $5\times 10^5\,T^\prime$ (the slower one associated with $k=5$). {With this eccentricity the behaviour of the arguments $\phi^{2:9}_1$, $\phi^{2:9}_3$ is less misleading regarding the importance of the short period librations discussed {with} the previous examples and their evolution is more clearly circulating. The reason is the small amplitude of the short period libration seen in $\phi^{2:9}_5$.}

The 2:9 resonance examples also illustrate how the inclination's relative variation is small and how the line of nodes of subpolar orbits regresses linearly at a rate that decreases as the inclination approaches $90^\circ$. {We also note} that when the eccentricity is increased, the {librating} critical argument's integer $k$ increases. This trend however is not generally valid as the next example will show.

\begin{figure*}
\begin{center}
{ 
\includegraphics[width=170mm]{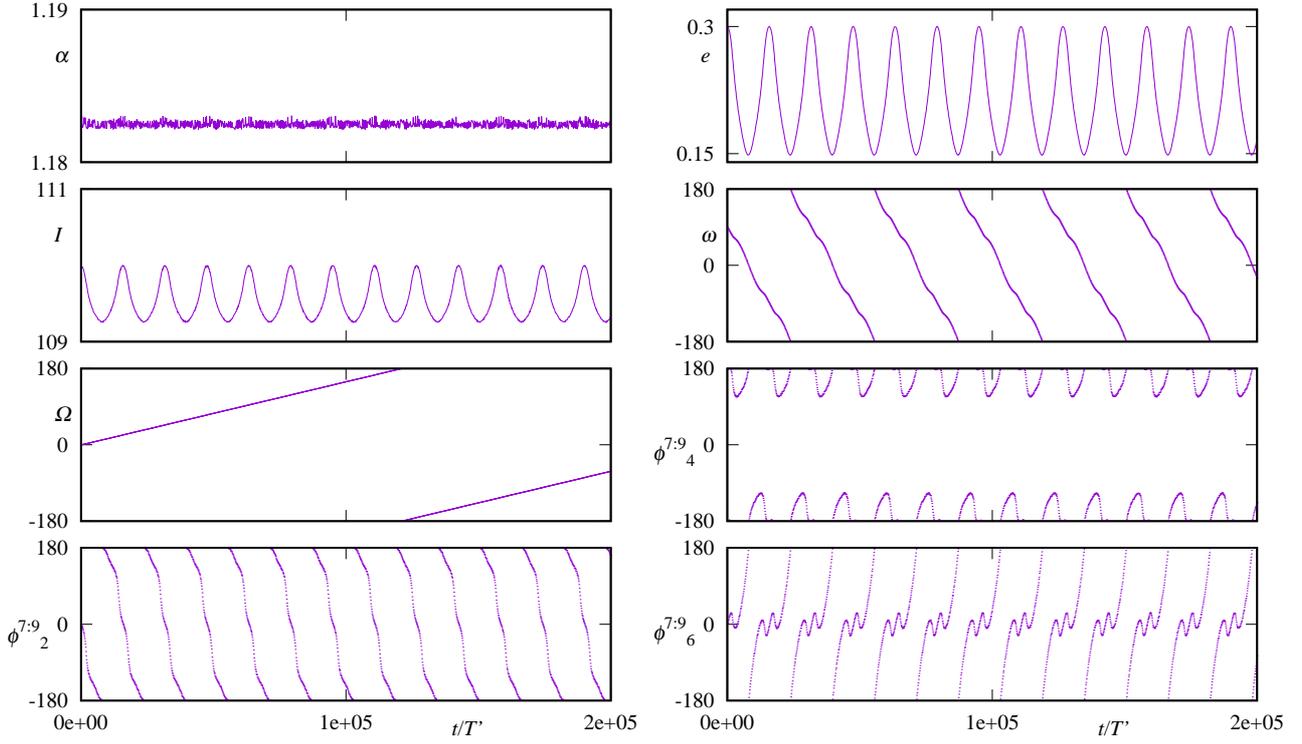}
}
\caption{Orbital elements and resonant angles' evolution of TNO 471325 at the outer 7:9 resonance with a Neptune mass planet in the context of the three-body problem. Initial parameters are:  eccentricity $e=0.3$, inclination $I=110^\circ$, longitude of ascending node $\Omega=0^\circ$, argument of pericentre $\omega=90^\circ$ and relative mean longitude $\lambda-\lambda^\prime= 180^\circ$.}\label{f7}
\end{center}
\end{figure*}

With the last example we examine the outer 7:9 resonance  in which  TNO 471325  \citep{Chen16} {could} currently be  captured. Since our interest is the polar disturbing function in the context of the three-body problem, these simulations do not reflect the actual evolution of the object but will give us an idea on the possible resonant arguments that might be involved. With this in mind, the initial orbital elements are: semimajor axis  $\alpha=1.182$,  eccentricity is $e=0.3$ and inclination $I=110^\circ$. We choose $\Omega=0^\circ$ and $\omega= 90^\circ$. As an even order resonance, the permissible resonant arguments are $\phi^{7:9}_{k}= 9\lambda-7\lambda^\prime-2\Omega-k\omega$ where $k$ is an even integer.  Figure \ref{f7} shows the evolution of the orbital elements as well as the arguments $\phi^{7:9}_{k}$ for $k=2, 4$ and $6$. It is seen that the argument $\phi^{7:9}_{4}$ for TNO 471325 in the three-body problem librates around $180^\circ$ with an amplitude and a period   of $68^\circ$ and $15550\,T^\prime$ respectively. {The argument of pericentre $\omega$ circulates rapidly, the longitude of ascending node $\Omega$ precesses linearly and the inclination has {moderate }variations as predicted in Section \ref{sec5}. The resonant arguments 
$\phi^{7:9}_{2}$ and $\phi^{7:9}_{6}$ both circulate but only the latter {displays} temporary librations similar to those of the previous resonance.}

\begin{figure*}
\begin{center}
{ 
\includegraphics[width=170mm]{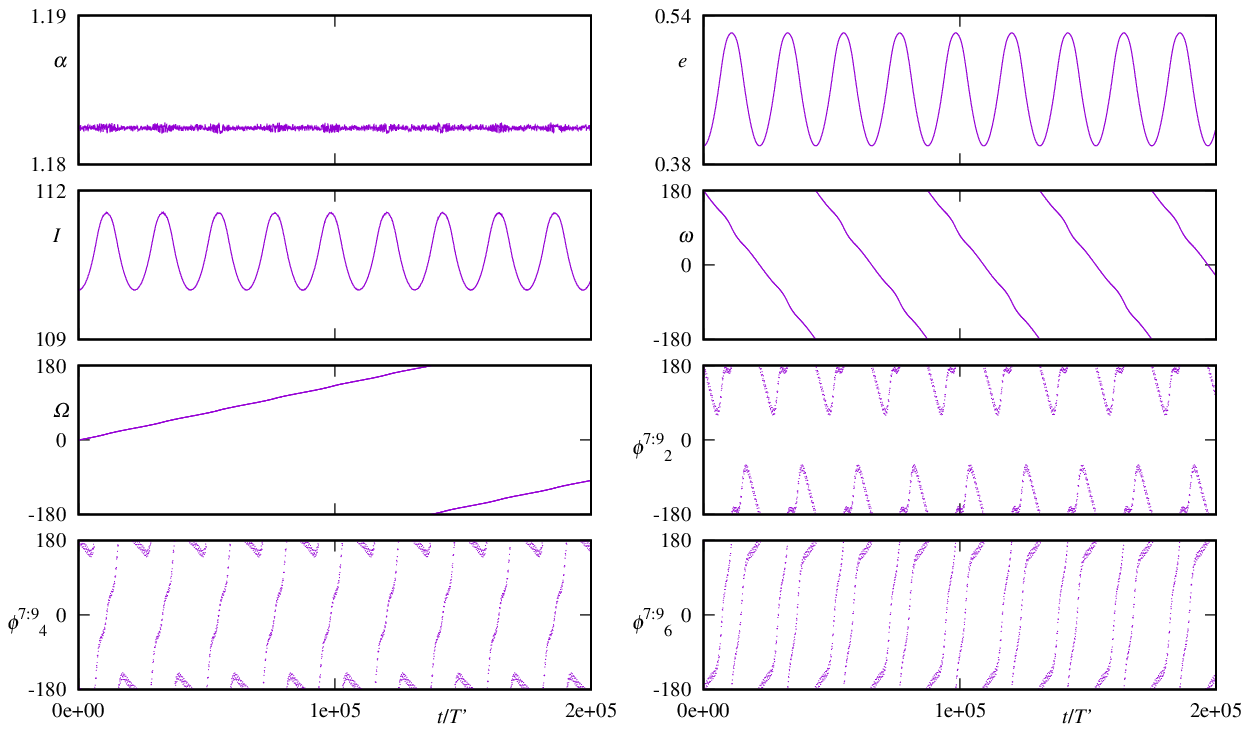}
}
\caption{Orbital elements and resonant angles' evolution of particle at the outer 7:9 resonance with a Neptune mass planet. Initial parameters are:  eccentricity $e=0.4$, inclination $I=110^\circ$, longitude of ascending node $\Omega=0^\circ$, argument of pericentre $\omega=180^\circ$ and relative mean longitude $\lambda-\lambda^\prime= 180^\circ$.}\label{f8}
\end{center}
\end{figure*}

Increasing the eccentricity to $e=0.4$ and initializing $\omega$ at $180^\circ$ makes the orbit librate with the critical argument $\phi^{7:9}_{2}$ with an amplitude of $120^\circ$ and a period of $22000\,T^\prime$ (Figure \ref{f8}). Thus the trend noted in the previous example, about how a larger $k$ could be associated with a larger $e$, is not confirmed. {The remaining  arguments  circulate but now it is $\phi^{7:9}_{4}$ that displays temporary librations whereas $\phi^{7:9}_{6}$ circulates with the fastest rate.}

\begin{figure*}
\begin{center}
{ 
\includegraphics[width=170mm]{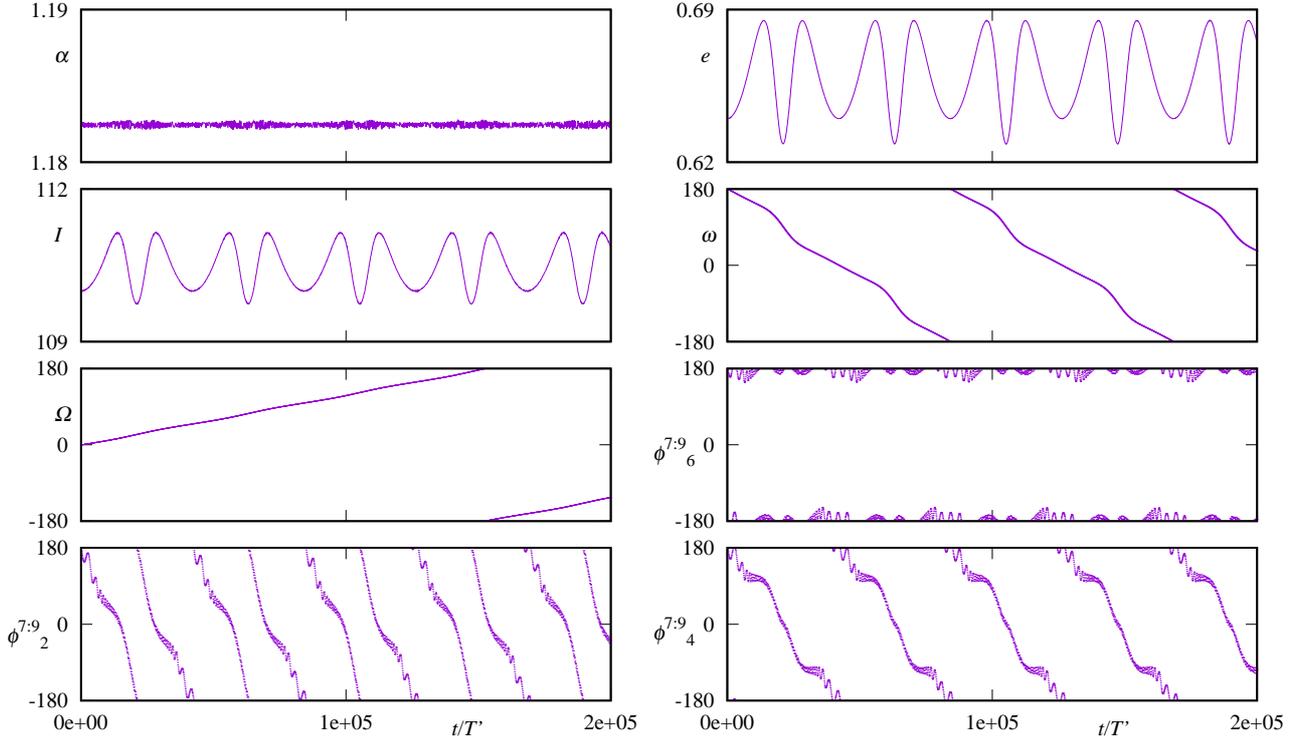}
}
\caption{Orbital elements and resonant angles' evolution of particle at the outer 7:9 resonance with a Neptune mass planet. Initial parameters are:  eccentricity $e=0.64$, inclination $I=110^\circ$, longitude of ascending node $\Omega=0^\circ$, argument of pericentre $\omega=180^\circ$ and relative mean longitude $\lambda-\lambda^\prime= 180^\circ$.}\label{f9}
\end{center}
\end{figure*}

 When the eccentricity is increased further to $e=0.64$ ($\omega=180^\circ$) ,the particle librates with the critical argument  $\phi^{7:9}_{6}$ of amplitude $70^\circ$ and period of $16000\,T^\prime$ (Figure \ref{f9}). {The arguments $\phi^{7:9}_{2}$ and $\phi^{7:9}_{4}$  circulate rapidly without temporary librations.}
 
We  conclude that for the same location and inclination as TNO 471325, the resonant argument depends strongly on the observed eccentricity. It is likely that  $\phi^{7:9}_{4}$ is the correct librating critical argument  as the effect of the other planets on resonant polar asteroids must be reduced because of their peculiar orbital geometry unless there is strong interaction  between the planets such as a mean motion resonance  that carries over to the motion of the polar asteroid.

\section{Concluding remarks}\label{sec7}
The main technical results of this work consist of (i) an explicit algorithm for generating a literal expansion of the disturbing function {for} nearly polar orbits of general order $N$ in eccentricity and inclination cosine (ii) the explicit form of the fourth order polar disturbing function through the direct part (\ref{RY}) with its force coefficients given in Tables \ref{t1} to \ref{t5}, the indirect part written explicitly in Table \ref{tind}, and the secular potential (\ref{SecPot}) whose force coefficients are given in Table \ref{tsec}.  Beyond the technical results, our original motivation for deriving a literal {expansion} of the disturbing function for nearly polar orbits is the realization that general {attitude} regarding resonance identification for polar Centaurs and TNOs is based on decades, and for some aspects more than a century, of use in planetary dynamics of the {classical} disturbing function derived for nearly coplanar prograde orbits.  It is therefore not surprising that we have {revealed} new features unseen in the {classical} disturbing function{\bf,} especially the structure of the force amplitudes that define resonance strength. In particular, the fact that regardless of resonance order, a particle {can} librate in the lowest harmonics (small $k$ in equation \ref{RY}) of the disturbing function is interesting as it explains an important observation that was made in our numerical studies of resonance capture at arbitrary inclination  \citep{NamouniMorais15,NamouniMorais17} that  resonance order is not a good indicator of resonance strength nor capture efficiency. This observation was particularly striking for the outer 1:5 resonance that exceeded 80 per cent capture efficiency for the most eccentric nearly polar orbits (Figure 6 of \cite{NamouniMorais17}). TNO 471325 provides a good example of a near polar asteroid locked in resonance. Our three-body simulations suggest that the resonant critical argument is $\phi_4^{7:9}$. Further simulations including all solar system planets are required to confirm this possibility and {discover} yet more Centaurs and TNOs in polar resonance with the giant planets.

\begin{table}
  \caption{Force coefficients $c^0_{mn}(p,q,\alpha)$ of the term $e^m\cos^n I \cos\phi$. }\label{t1}
   \begin{tabular}{cl}
    \hline
$c_{00}^0 $&$\frac{1}{2}A_{0, p, q, 0},$\\
$c_{01}^0 $&$ \frac{\alpha}{8} (A_{1, p-1, q-1, 0} - A_{1, p-1, q+1, 0} -  A_{1, p+1, q-1, 0} $\nonumber\\&$+ A_{1, p+1, q+1, 0}) ,$\\
$c_{20}^0 $&$ \frac{1}{8} (-4 q^2 A_{0, p, q, 0} + 2 A_{0, p, q, 1} +  A_{0, p, q, 2}),$\\
$c_{02}^0 $&$\frac{3\alpha^2}{64} (A_{2, p-2, q-2, 0} - 2 A_{2, p-2, q, 0} + 
A_{2, p-2, q+2, 0}$\nonumber\\&$ - 2 A_{2, p, q-2, 0}+ 4 A_{2, p, q, 0} - 
2 A_{2, p, q+2, 0} $\nonumber\\&$+ A_{2, p+2, q-2, 0} - 
2 A_{2, p+2, q, 0} + A_{2, p+2, q+2, 0})$\\
$c_{21}^0 $&$-\frac{\alpha}{32} [( 4 q^2-2) (A_{1, p-1, q-1, 0}-A_{1, p-1, q+1, 0}$\nonumber\\&$-A_{1, p+1, q-1, 0}+A_{1, p+1, q+1, 0} ) - 4 (A_{1, p-1, q-1, 1}$\nonumber\\&$+A_{1, p-1, q+1, 1}+A_{1, p+1, q-1, 1} -A_{1, p+1, q+1, 1}) $\nonumber\\&$- A_{1, p-1, q-1, 2} + A_{1, p-1, q+1, 2} + A_{1, p+1, q-1, 2}   $\nonumber\\&$- A_{1, p+1, q+1, 2}]$\\
$c_{03}^0 $&$\frac{5\alpha^3}{256} [ 3 (A_{3, p-3, q+1, 0}  - 
    A_{3, p-3, q-1, 0}  -  A_{3, p-1, q-3, 0}$\nonumber\\&$ + 
    A_{3, p-1, q+3, 0} +  A_{3, p+1, q-3, 0}  - 
    A_{3, p+1, q+3, 0} $\nonumber\\&$ + 
    A_{3, p+3, q-1, 0} - A_{3, p+3, q+1, 0}) +
   9( A_{3, p-1, q-1, 0} $\nonumber\\&$-  A_{3, p-1, q+1, 0} - 
    A_{3, p+1, q-1, 0} +  A_{3, p+1, q+1, 0}) $\nonumber\\&$- A_{3, p+3, q-3, 0}- 
   A_{3, p-3, q+3, 0}+    A_{3, p-3, q-3, 0}$\nonumber\\&$+A_{3, p+3, q+3, 0}]$\\
$c_{40}^0 $&$\frac{1}{128} [q^2 (16 q^2-9) A_{0, p, q, 0} - 8 q^2 (A_{0, p, q, 1} +
    A_{0, p, q, 2}) $\nonumber\\&$+ 4 A_{0, p, q, 3} + A_{0, p, q, 4}]$\\
$c_{04}^0 $&$\frac{35\alpha^4}{4096} [A_{4, p-4, q+4, 0}+A_{4, p-4, q-4, 0} + 
   A_{4, p+4, q-4, 0}  $\nonumber\\&$+ 
   A_{4, p+4, q+4, 0} 
   -4 (A_{4, p-4, q-2, 0}   + 
    A_{4, p-4, q+2, 0} $\nonumber\\&$+
    A_{4, p-2, q-4, 0}  + 
   A_{4, p-2, q+4, 0}+  A_{4, p+2, q-4, 0} +$\nonumber\\&$+ A_{4, p+2, q+4, 0}+ A_{4, p+4, q-2, 0} + A_{4, p+4, q+2, 0} )$\nonumber\\&$
   + 6 (A_{4, p-4, q, 0}+  A_{4, p, q-4, 0}+ 
    A_{4, p, q+4, 0}$\nonumber\\&$+ 
    A_{4, p+4, q, 0})
   + 16 (A_{4, p-2, q-2, 0} + A_{4, p-2, q+2, 0}$\nonumber\\&$+ 
    A_{4, p+2, q+2, 0} + 
    A_{4, p+2, q-2, 0}) - 
   24 (A_{4, p-2, q, 0}$\nonumber\\&$ - 
   A_{4, p, q-2, 0}  -  A_{4, p, q+2, 0}   - A_{4, p+2, q, 0}) $\nonumber\\&$+ 36 A_{4, p, q, 0} ]$\\
   $c_{22}^0 $&$-\frac{3\alpha^2}{256} [2( 2 q^2-3) (A_{2, p-2, q-2, 0}-2 A_{2, p-2, q, 0}$\nonumber\\&$+ A_{2, p-2, q+2, 0} - 2 A_{2, p, q-2, 0}
  - 2 A_{2, p, q+2, 0} $\nonumber\\&$+    4A_{2, p, q, 0}-2 A_{2, p+2, q, 0} +A_{2, p+2, q-2, 0}$\nonumber\\&$+A_{2, p+2, q+2, 0}) - 
   6 A_{2, p-2, q-2, 1} - A_{2, p-2, q-2, 2}  $\nonumber\\&$+ 
   12 A_{2, p-2, q, 1} + 2 A_{2, p-2, q, 2}  - 
   6 A_{2, p-2, q+2, 1} $\nonumber\\&$- A_{2, p-2, q+2, 2}  + 
   12 A_{2, p, q-2, 1} + 2 A_{2, p, q-2, 2}  $\nonumber\\&$- 24 A_{2, p, q, 1} - 4 A_{2, p, q, 2} + 
   12 A_{2, p, q+2, 1} $\nonumber\\&$+ 2 A_{2, p, q+2, 2}  - 
   6 A_{2, p+2, q-2, 1} - A_{2, p+2, q-2, 2} $\nonumber\\&$ + 
   12 A_{2, p+2, q, 1} + 2 A_{2, p+2, q, 2} - 
   6 A_{2, p+2, q+2, 1}$\nonumber\\&$ - A_{2, p+2, q+2, 2}]$\\
 \hline
  \end{tabular}
 \end{table}

\begin{table}
  \caption{Force coefficients $c^{1}_{mn}(p,q,\alpha)$ of the term $e^m\cos^n I \cos(\phi-\omega)$. }\label{t2}
   \begin{tabular}{cl}    \hline
$c_{10}^{1} $&$-\frac{1}{4} [2 (1 - q) A_{0, p, q-1, 0} + A_{0, p, q-1, 1}],$\\ 
$c_{11}^{1} $&$ \frac{\alpha}{16}  [( 2 q-3) (A_{1, p-1, q-2, 0} - A_{1, p-1, q, 0}+ A_{1, p+1, q, 0}$\nonumber\\&$- A_{1, p+1, q-2, 0}) - A_{1, p-1, q-2, 1}  + A_{1, p-1, q, 1} $\nonumber\\&$ +  A_{1, p+1, q-2, 1}  - A_{1, p+1, q, 1}],$\\
$c_{12}^{1} $&$-\frac{3\alpha^2}{128} [2 (2-q) (A_{2, p-2, q-3, 0}-2A_{2, p-2, q-1, 0} $\nonumber\\&$+A_{2, p-2, q+1, 0} 
                              -2A_{2, p, q-3, 0}+4A_{2, p, q-1, 0}$\nonumber\\&$+ A_{2, p+2, q-3, 0}-2A_{2, p, q+1, 0}-2 A_{2, p+2, q-1, 0}$\nonumber\\&$+A_{2, p+2, q+1, 0} ) + 
   A_{2, p-2, q-3, 1} 
    - 2 A_{2, p-2, q-1, 1} $\nonumber\\&$
    + A_{2, p-2, q+1, 1} 
   - 2 A_{2, p, q-3, 1} 
     + 4 A_{2, p, q-1, 1}$\nonumber\\&$
       - 2 A_{2, p, q+1, 1} 
       + A_{2, p+2, q-3, 1} 
        - 2 A_{2, p+2, q-1, 1} $\nonumber\\&$
       + A_{2, p+2, q+1, 1}]$\\
$c_{30}^1 $&$\frac{1}{32} [2 q (7 q - 4 q^2-3) A_{0, p, q-1, 0} + 
   q (4 q-1) A_{0, p, q-1, 1} $\nonumber\\&$- 4 A_{0, p, q-1, 2} + 
   2 q A_{0, p, q-1, 2} - A_{0, p, q-1, 3}]$\\
 $c_{13}^1 $&$\frac{5\alpha^3}{512} [(2 q-5) (A_{3, p-3, q-4, 0} -3A_{3, p-3, q-2, 0} $\nonumber\\&$+3 A_{3, p-3, q, 0}-A_{3, p-3, q+2, 0}-3A_{3, p-1, q-4, 0} $\nonumber\\&$+9A_{3, p-1, q-2, 0} -9A_{3, p-1, q, 0} +3A_{3, p-1, q+2, 0} $\nonumber\\&$+3A_{3, p+1, q-4, 0} -9A_{3, p+1, q-2, 0}+9 A_{3, p+1, q, 0}$\nonumber\\&$-3A_{3, p+1, q+2, 0}-A_{3, p+3, q-4, 0} +3A_{3, p+3, q-2, 0}$\nonumber\\&$-3A_{3, p+3, q, 0} +A_{3, p+3, q+2, 0})
 -   A_{3, p-3, q-4, 1} $\nonumber\\&$+  A_{3, p+3, q-4, 1} - A_{3, p+3, q+2, 1} + A_{3, p-3, q+2, 1} $\nonumber\\&$
   + 3 (A_{3, p-3, q-2, 1}
      -  A_{3, p-3, q, 1}     
    +  A_{3, p-1, q-4, 1}   $\nonumber\\&$  
    -  A_{3, p-1, q+2, 1} 
     -  A_{3, p+1, q-4, 1} 
          +  A_{3, p+1, q+2, 1}   $\nonumber\\&$   
      -  A_{3, p+3, q-2, 1} 
    +   A_{3, p+3, q, 1} )
    + 9 (
     A_{3, p-1, q, 1}$\nonumber\\&$ -A_{3, p-1, q-2, 1}
   +  A_{3, p+1, q-2, 1} 
        -  A_{3, p+1, q, 1})] $\\
     $c_{31}^1 $&   $-\frac{\alpha}{128}  [q (7 - 18 q + 8 q^2) (A_{1, p-1, q-2, 0} - A_{1, p-1, q, 0}$\nonumber\\&$ -A_{1, p+1, q-2, 0} +A_{1, p+1, q, 0} )$\nonumber\\&$ 
     + (8 - 3 q - 4 q^2) (A_{1, p-1, q-2, 1} -A_{1, p-1, q, 1}$\nonumber\\&$ -A_{1, p+1, q-2, 1}+A_{1, p+1, q, 1} )
     + (7-2q) (A_{1, p-1, q-2, 2}$\nonumber\\&$ - A_{1, p-1, q, 2} -A_{1, p+1, q-2, 2}+A_{1, p+1, q, 2})$\nonumber\\&$ 
     +  A_{1, p-1, q-2, 3}  - A_{1, p-1, q, 3}         -    A_{1, p+1, q-2, 3} $\nonumber\\&$ 
        + A_{1, p+1, q, 3}]$\\
 \hline
  \end{tabular}
 \end{table}

\begin{table}
  \caption{Force coefficients $c^{2}_{mn}(p,q,\alpha)$ of the term $e^m\cos^n I \cos(\phi-2\omega)$. }\label{t3}
   \begin{tabular}{cl}
    \hline
    $c^{2}_{20} $&$\frac{1}{16}[(6 - 11 q +   4 q^2) A_{0, p, q-2, 0} + (6  -  4 q) A_{0, p, q-2, 1} $\nonumber\\&$+ A_{0, p, q-2, 2}],$\\ 
    $c_{21}^2 $&$\frac{\alpha}{64}  [(4 q^2-15q+12) (A_{1, p-1, q-3, 0}-A_{1, p-1, q-1, 0}$\nonumber\\&$-A_{1, p+1, q-3, 0}+A_{1, p+1, q-1, 0}) 
    $\nonumber\\&$-    4 (q-2) (A_{1, p-1, q-3, 1}+A_{1, p-1, q-1, 1}$\nonumber\\&$+A_{1, p+1, q-3, 1}-A_{1, p+1, q-1, 1}) + A_{1, p-1, q-3, 2}$\nonumber\\&$ - A_{1, p-1, q-1, 2} 
    - A_{1, p+1, q-3, 2} + A_{1, p+1, q-1, 2}]$\\
    $c_{40}^2 $&$\frac{1}{192} [(12 + 26 q - 88 q^2 + 68 q^3 - 16 q^4) A_{0, p, q-2, 0} $\nonumber\\&$- 
   2 (6 - 23 q +24 q^2 - 8 q^3) A_{0, p, q-2, 1} $\nonumber\\&$+ 
   (6 -9q) A_{0, p, q-2, 2}  + 
   4(2-q) A_{0, p, q-2, 3}$\nonumber\\&$+ A_{0, p, q-2, 4}]$\\
     $c_{22}^2 $&$\frac{3\alpha^2}{512} [(20 - 19 q + 4 q^2) (A_{2, p-2, q-4,   0} $\nonumber\\
     &$-2A_{2, p-2, q-2, 0} +A_{2, p-2, q, 0}-2A_{2, p, q-4, 0}$\nonumber\\
     &$+4 A_{2, p, q-2, 0}-2A_{2, p, q, 0} +A_{2, p+2, q-4, 0}$\nonumber\\&$-2A_{2, p+2, q-2, 0}+A_{2, p+2, q, 0})$\nonumber\\&$
     + (10 - 4 q) (A_{2, p-2, q-4, 1}-2A_{2, p-2, q-2, 1} +A_{2, p-2, q, 1}$\nonumber\\&$-2A_{2, p, q-4, 1} +4A_{2, p, q-2, 1} -2A_{2, p, q, 1}$\nonumber\\&$+A_{2, p+2, q-4, 1}-2A_{2, p+2, q-2, 1}+A_{2, p+2, q, 1}) $\nonumber\\&$
     + A_{2, p-2, q-4, 2} 
     - 2 A_{2, p-2, q-2, 2} 
     + A_{2, p-2, q, 2}$\nonumber\\&$
     - 2 A_{2, p, q-4, 2} 
     + 4 A_{2, p, q-2, 2} 
     -  2 A_{2, p, q, 2} $\nonumber\\&$
      +A_{2, p+2, q-4, 2}
     -2 A_{2, p+2, q-2, 2} 
  + A_{2, p+2, q, 2}]$\\
 \hline
  \end{tabular}
 \end{table}

\begin{table}
  \caption{Force coefficients $c^{3}_{mn}(p,q,\alpha)$ of the term $e^m\cos^n I \cos(\phi-3\omega)$. }\label{t4}
   \begin{tabular}{cl}
    \hline
    $c^{3}_{30} $&$\frac{1}{96} [(8 q^3- 42 q^2+ 62 q -24 ) A_{0, p, q-3, 0} $\nonumber\\&$- 
   3 (12 - 15 q + 4 q^2) A_{0, p, q-3, 1} +(6q- 12) A_{0, p, q-3, 2} $\nonumber\\&$ - A_{0, p, q-3, 3}]$\\ 
   $c_{31}^3 $&   $\frac{\alpha}{384} [(60 - 107 q + 54 q^2 - 8 q^3) (A_{1, p-1, q-2, 0} -A_{1, p-1, q-4, 0}$\nonumber\\&$+A_{1, p+1, q-4, 0}-A_{1, p+1, q-2, 0}   )$\nonumber\\&$
   - 3 (20 - 19 q + 4 q^2) (A_{1, p-1, q-4, 1}- A_{1, p-1, q-2, 1}$\nonumber\\&$ -A_{1, p+1, q-4, 1}+A_{1, p+1, q-2, 1} ) 
   + (15-6q) (A_{1, p-1, q-2, 2} $\nonumber\\&$-A_{1, p-1, q-4, 2}+A_{1, p+1, q-4, 2} -A_{1, p+1, q-2, 2})$\nonumber\\&$
   - A_{1, p-1, q-4, 3} 
   + A_{1, p-1, q-2, 3} 
   + A_{1, p+1, q-4, 3} $\nonumber\\&$
   -  A_{1, p+1, q-2, 3}]$\\
 \hline
  \end{tabular}
 \end{table}

\begin{table}
  \caption{Force coefficients $c^{4}_{mn}(p,q,\alpha)$ of the term $e^m\cos^n I \cos(\phi-4\omega)$. }\label{t5}
   \begin{tabular}{cl}
    \hline
    $c^{4}_{40} $&$\frac{1}{768} [(120 - 394 q + 379 q^2 - 136 q^3 + 16 q^4) A_{0, p, q-4, 
     0} $\nonumber\\&$- 4 (-60 + 107 q - 54 q^2 + 8 q^3) A_{0, p, q-4, 1} $\nonumber\\&$+ 
   (120-114q+24q^2) A_{0, p, q-4, 2}$\nonumber\\&$+ (20-8q) A_{0, p, q-4, 3}  + A_{0, p, q-4, 4}]$\\ 
 \hline
  \end{tabular}
 \end{table}

\begin{table}
  \caption{Force amplitudes and cosine arguments of  the indirect part. }\label{tind}

   \begin{tabular}{ll}    \hline
     Cosine argument & Force amplitude\\ \hline\\
        $ \lambda^\prime-\varpi$&$\frac{3\alpha}{4} (1 + {\cos I}) e $\\ 
     $\lambda - \lambda^\prime$&$-\frac{\alpha}{2} (1+ {\cos I}  - \frac{1}{2}e^2  -
 \frac{1}{2} e^2{\cos I}  - \frac{1}{64}e^4 $)\\   
    $\lambda - \lambda^\prime- 2 \varpi + 2 \Omega$&$-\frac{\alpha}{48} e^2 (3 +e^2- 3 {\cos I}) $\\ 
   $2 \lambda - \lambda^\prime- \varpi $&$\frac{\alpha}{16} (1 + {\cos I})  (3 e^3-4e)  $\\ 
       $2 \lambda - \lambda^\prime- 3 \varpi + 2 \Omega $&$\frac{\alpha}{48} ( {\cos I}-1) e^3 $\\ 
      $3 \lambda - \lambda^\prime- 2 \varpi $&$\frac{3\alpha}{16} e^2 (e^2- {\cos I} -1) $\\ 
          $3 \lambda - \lambda^\prime- 4 \varpi + 2 \Omega $&$-\frac{3\alpha}{256} e^4 $\\ 
     $4 \lambda - \lambda^\prime- 3 \varpi $&$-\frac{\alpha}{6}(1+  {\cos I}) e^3 $\\   
     $5 \lambda - \lambda^\prime- 4 \varpi $&$-\frac{125\alpha}{768} e^4 $\\ 
     $ \lambda^\prime+\varpi - 2 \Omega $&$-\frac{3\alpha}{4} ({\cos I}-1) e $\\ 
      $\lambda +\lambda^\prime- 2 \Omega $&$-\frac{\alpha}{128} (64 - 32 e^2 - e^4 +32  (e^2-2)\cos I) $\\ 
    $\lambda + \lambda^\prime- 2 \varpi $&$-\frac{\alpha}{48} e^2 (3 + e^2+3 {\cos I} $\\ 
     $2 \lambda + \lambda^\prime- 3 \varpi $&$-\frac{\alpha}{48} (1 + {\cos I}) e^3 $\\ 
    $2 \lambda + \lambda^\prime- \varpi - 2 \Omega$&$-\frac{\alpha}{16} ( {\cos I}-1) e (3 e^2-4) $\\ 
        $3 \lambda + \lambda^\prime- 4 \varpi $&$-\frac{3\alpha}{256} e^4 $\\   
 $3 \lambda + \lambda^\prime- 2 \varpi -2 \Omega $&$-\frac{3\alpha}{16} e^2 (1 -e^2- {\cos I} )$\\
      $4 \lambda + \lambda^\prime- 3 \varpi - 2 \Omega $&$\frac{\alpha}{6} ( {\cos I}-1) e^3 $\\ 
     $5 \lambda + \lambda^\prime- 4 \varpi - 2 \Omega $&$-\frac{125\alpha}{768} e^4 $\\ 
 \hline
  \end{tabular}
 \end{table}

\begin{table}  \caption{Force coefficients $s^{k}_{mn}(\alpha)$ of the secular term $e^{m}\cos^{n} I\cos (k\omega)$. }\label{tsec}
   \begin{tabular}{ccl}
    \hline
   $k=0$  & $s^{0}_{20} $ &$\frac{1}{16}(2 A_{0, 0, 0, 1} + A_{0, 0, 0, 2})$\\ 
               & $s^{0}_{02} $ &$\frac{3\alpha^2}{32}(A_{2, 0, 0, 0} - 2 A_{2, 2, 0, 0} +     A_{2, 2, 2, 0})$\\ 
                  & $s^{0}_{22} $&$\frac{3\alpha^2}{128}  (6 A_{2, 0, 0, 0} + 6 A_{2, 0, 0, 1} + 
     A_{2, 0, 0, 2}$\nonumber\\&&$ - 12 A_{2, 2, 0, 0} - 12 A_{2, 2, 0, 1} - 
     2 A_{2, 2, 0, 2} $\nonumber\\&&$+ 6 A_{2, 2, 2, 0} + 6 A_{2, 2, 2, 1} + 
     A_{2, 2, 2, 2})$\\ 
      &$s^{0}_{40} $&$\frac{1}{256} (4 A_{0, 0, 0, 3} + A_{0, 0, 0, 4})$\\ 
      &$s^{0}_{04} $&$\frac{35\alpha^4}{2048}  (9 A_{4, 0, 0, 0} - 24 A_{4, 0, 2, 0} + 
   3 A_{4, 0, 4, 0} $\nonumber\\&&$+ 16 A_{4, 2, 2, 0} + 3 A_{4, 4, 0, 0} - 
   8 A_{4, 4, 2, 0} $\nonumber\\&&$+ A_{4, 4, 4, 0})$\\
    $k=2$&$s^{2}_{20} $&$\frac{1}{16}(6 A_{0, 0, 2, 0} + 6 A_{0, 0, 2, 1} + A_{0, 0, 2, 2})$\\
     &$s^{2}_{22} $&$-\frac{3\alpha^2}{256} (20 A_{2, 0, 0, 0} + 10 A_{2, 0, 0, 1} + A_{2, 0, 0, 2}$\nonumber\\&&$ - 
  60 A_{2, 2, 0, 0} - 30 A_{2, 2, 0, 1} - 3 A_{2, 2, 0, 2} $\nonumber\\&&$+ 
  40 A_{2, 2, 2, 0} + 20 A_{2, 2, 2, 1} + 2 A_{2, 2, 2, 2}$\nonumber\\&&$ - 
  20 A_{2, 2, 4, 0} - 10 A_{2, 2, 4, 1} - A_{2, 2, 4, 2} $\nonumber\\&&$+ 
  20 A_{2, 4, 0, 0} + 10 A_{2, 4, 0, 1} + A_{2, 4, 0, 2})$\\
   &$s^{2}_{40} $&$\frac{1}{192} (12 A_{0, 0, 2, 0} - 12 A_{0, 0, 2, 1} + 6 A_{0, 0, 2, 2} $\nonumber\\&&$+ 
   8 A_{0, 0, 2, 3} + A_{0, 0, 2, 4})$\\ 
    $k=4$&$s^{4}_{40} $&$\frac{1}{768} (120 A_{0, 0, 4, 0} + 240 A_{0, 0, 4, 1}$\nonumber\\&&$ + 120 A_{0, 0, 4, 2} + 
   20 A_{0, 0, 4, 3} + A_{0, 0, 4, 4})$\\
 \hline
  \end{tabular}
 \end{table}

\newpage

\section*{Acknowledgments}
The authors thank Zoran Kne{\v{z}}evi\'c for his thorough and helpful review of the manuscript. F.N. thanks the University of Rio Claro (UNESP) for their hospitality during his stay when part of this work was performed. This work was supported by grant 2015/17962-5 of S\~ao Paulo Research Foundation (FAPESP).
\bibliographystyle{mnras}

\bibliography{ms}

\begin{thebibliography}{}
\makeatletter
\relax
\def\mn@urlcharsother{\let\do\@makeother \do\$\do\&\do\#\do\^\do\_\do\%\do\~}
\def\mn@doi{\begingroup\mn@urlcharsother \@ifnextchar [ {\mn@doi@}
  {\mn@doi@[]}}
\def\mn@doi@[#1]#2{\def\@tempa{#1}\ifx\@tempa\@empty \href
  {http://dx.doi.org/#2} {doi:#2}\else \href {http://dx.doi.org/#2} {#1}\fi
  \endgroup}
\def\mn@eprint#1#2{\mn@eprint@#1:#2::\@nil}
\def\mn@eprint@arXiv#1{\href {http://arxiv.org/abs/#1} {{\tt arXiv:#1}}}
\def\mn@eprint@dblp#1{\href {http://dblp.uni-trier.de/rec/bibtex/#1.xml}
  {dblp:#1}}
\def\mn@eprint@#1:#2:#3:#4\@nil{\def\@tempa {#1}\def\@tempb {#2}\def\@tempc
  {#3}\ifx \@tempc \@empty \let \@tempc \@tempb \let \@tempb \@tempa \fi \ifx
  \@tempb \@empty \def\@tempb {arXiv}\fi \@ifundefined
  {mn@eprint@\@tempb}{\@tempb:\@tempc}{\expandafter \expandafter \csname
  mn@eprint@\@tempb\endcsname \expandafter{\@tempc}}}

\bibitem[\protect\citeauthoryear{{Brouwer} \& {Clemence}}{{Brouwer} \&
  {Clemence}}{1961}]{BC61}
{Brouwer} D.,  {Clemence} G.~M.,  1961, {Celestial Mechanics}.
{Academic Press, New York}

\bibitem[\protect\citeauthoryear{{Chen} et~al.,}{{Chen} et~al.}{2016}]{Chen16}
{Chen} Y.-T.,  et~al., 2016, \mn@doi [\apjl] {10.3847/2041-8205/827/2/L24},
  \href {http://adsabs.harvard.edu/abs/2016ApJ...827L..24C} {827, L24}

\bibitem[\protect\citeauthoryear{{Ellis} \& {Murray}}{{Ellis} \&
  {Murray}}{2000}]{EllisMurray00}
{Ellis} K.~M.,  {Murray} C.~D.,  2000, \mn@doi [\icarus]
  {10.1006/icar.2000.6399}, \href
  {http://adsabs.harvard.edu/abs/2000Icar..147..129E} {147, 129}

\bibitem[\protect\citeauthoryear{{Gladman} et~al.,}{{Gladman}
  et~al.}{2009}]{Gladmanetal09}
{Gladman} B.,  et~al., 2009, \mn@doi [\apjl] {10.1088/0004-637X/697/2/L91},
  \href {http://adsabs.harvard.edu/abs/2009ApJ...697L..91G} {697, L91}

\bibitem[\protect\citeauthoryear{{Hagihara}}{{Hagihara}}{1972}]{Hagihara}
{Hagihara} Y.,  1972, {Celestial mechanics. Vol.2, Perturbation theory}.
Massachusetts Institute of Technology (MIT), Cambridge, MS

\bibitem[\protect\citeauthoryear{{Kozai}}{{Kozai}}{1962}]{Kozai62}
{Kozai} Y.,  1962, \mn@doi [\aj] {10.1086/108790}, \href
  {http://adsabs.harvard.edu/abs/1962AJ.....67..591K} {67, 591}

\bibitem[\protect\citeauthoryear{{Laskar} \& {Bou{\'e}}}{{Laskar} \&
  {Bou{\'e}}}{2010}]{Laskar10}
{Laskar} J.,  {Bou{\'e}} G.,  2010, \mn@doi [\aap]
  {10.1051/0004-6361/201014496}, \href
  {http://adsabs.harvard.edu/abs/2010A%26A...522A..60L} {522, A60}

\bibitem[\protect\citeauthoryear{{Lidov}}{{Lidov}}{1962}]{Lidov62}
{Lidov} M.~L.,  1962, \mn@doi [\planss] {10.1016/0032-0633(62)90129-0}, \href
  {http://adsabs.harvard.edu/abs/1962P%26SS....9..719L} {9, 719}

\bibitem[\protect\citeauthoryear{{Morais} \& {Giuppone}}{{Morais} \&
  {Giuppone}}{2012}]{MoraisGiuppone12}
{Morais} M.~H.~M.,  {Giuppone} C.~A.,  2012, \mn@doi [\mnras]
  {10.1111/j.1365-2966.2012.21151.x}, \href
  {http://adsabs.harvard.edu/abs/2012MNRAS.424...52M} {424, 52}

\bibitem[\protect\citeauthoryear{{Morais} \& {Namouni}}{{Morais} \&
  {Namouni}}{2013a}]{MoraisNamouni13a}
{Morais} M.~H.~M.,  {Namouni} F.,  2013a, \mn@doi [Celestial Mechanics and
  Dynamical Astronomy] {10.1007/s10569-013-9519-2}, \href
  {http://adsabs.harvard.edu/abs/2013CeMDA.117..405M} {117, 405}

\bibitem[\protect\citeauthoryear{{Morais} \& {Namouni}}{{Morais} \&
  {Namouni}}{2013b}]{MoraisNamouni13b}
{Morais} M.~H.~M.,  {Namouni} F.,  2013b, \mn@doi [\mnras]
  {10.1093/mnrasl/slt106}, \href
  {http://adsabs.harvard.edu/abs/2013MNRAS.436L..30M} {436, L30}

\bibitem[\protect\citeauthoryear{{Murray} \& {Dermott}}{{Murray} \&
  {Dermott}}{1999}]{ssdbook}
{Murray} C.~D.,  {Dermott} S.~F.,  1999, {Solar system dynamics}.
{Cambridge University Press}

\bibitem[\protect\citeauthoryear{{Namouni} \& {Morais}}{{Namouni} \&
  {Morais}}{2015}]{NamouniMorais15}
{Namouni} F.,  {Morais} M.~H.~M.,  2015, \mn@doi [\mnras]
  {10.1093/mnras/stu2199}, \href
  {http://adsabs.harvard.edu/abs/2015MNRAS.446.1998N} {446, 1998}

\bibitem[\protect\citeauthoryear{{Namouni} \& {Morais}}{{Namouni} \&
  {Morais}}{2017}]{NamouniMorais17}
{Namouni} F.,  {Morais} M.~H.~M.,  2017, \mn@doi [\mnras]
  {10.1093/mnras/stx290}, 467, 2673

\bibitem[\protect\citeauthoryear{{Quinn}, {Tremaine}  \& {Duncan}}{{Quinn}
  et~al.}{1990}]{Quinn90}
{Quinn} T.,  {Tremaine} S.,   {Duncan} M.,  1990, \mn@doi [\apj]
  {10.1086/168800}, \href {http://adsabs.harvard.edu/abs/1990ApJ...355..667Q}
  {355, 667}

\bibitem[\protect\citeauthoryear{{Williams}}{{Williams}}{1969}]{Williams69}
{Williams} J.~G.,  1969, PhD thesis, University of California at Los Angeles.

\makeatother
\end{thebibliography}

\end{document}